\documentclass[useAMS,usenatbib]{mn2e}

\usepackage{graphics,mssymb}

\title[The corona and upper transition region of $\epsilon$\,Eridani]
{The corona and upper transition region of $\epsilon$\,Eridani}
\author[J.-U. Ness and C. Jordan]{J.-U. Ness$^{1,2}$\thanks{E-mail:
Jan-Uwe.Ness@asu.edu} and C. Jordan$^{1}$\\
$^1$Department of Physics, Rudolf Peierls Centre for Theoretical Physics,
University of Oxford, 1 Keble Road, Oxford OX1\,3NP, UK\\
$^2$School of Earth and Space Exploration, Arizona
State University, Tempe, AZ 85287-1404, USA}

\begin{document}

\date{Accepted ; Received \today}

\pagerange{\pageref{firstpage}--\pageref{lastpage}} \pubyear{2006}

\maketitle

\label{firstpage}

\begin{abstract}
We present analyses of observations of $\epsilon$\,Eridani (K2~V)
made with the Low Energy Transmission Grating Spectrometer on {\it Chandra}
and the {\it Extreme Ultraviolet Explorer}, supplemented by observations made
 with the Space Telescope Imaging Spectrograph, the {\it Far Ultraviolet 
Spectroscopic Explorer} and the Reflection Grating
Spectrometer on {\it XMM-Newton}. The observed emission lines are used to
find relative element abundances, to place limits on the electron densities
and pressures and to determine the mean apparent emission measure distribution.
As in the previous paper by \cite{simjordan03a}, the mean emitting area as a
function of the electron temperature is derived by comparisons with a
theoretical emission measure distribution found from energy balance 
arguments. The final model has a coronal temperature of $3.4 \times 10^6$~K,
an electron pressure of $1.3 \times 10^{16}$~cm$^{-3}$K at $T_{\rm e}
= 2 \times 10^5$~K and an area 
filling factor of 0.14 at $3.2 \times 10^5$~K. We discuss a number of issues 
concerning the atomic data currently available. Our analyses are based mainly
 on the latest version of CHIANTI (v5.2). We conclude that the Ne/O relative 
abundance is 0.30, larger than that recommended from solar studies, and that 
there is no convincing evidence for enhanced coronal abundances of elements 
with low first ionization potentials.
\end{abstract}


\section{Introduction}
\label{S1}
$\epsilon$~Eri (K2~V) is a nearby dwarf star that has been observed
over a wide spectral range, from the infrared to X-ray wavelengths. 
Its fundamental parameters have been discussed by \cite{drakesmith93}, 
where references to earlier work can be found, and more recently by
\cite{difolco04} and \cite{ap04}. \cite{difolco04} have derived the 
diameter of $\epsilon$~
Eri from interferometric measurements and have also made models to check the
self-consistency of their adopted parameters. A more recent measurement by
\cite{difolco07} gives a radius that is smaller by only 
0.008~$R_{\odot}$. The values of the parameters
adopted here are given in Table~\ref{tab1} and are from \cite{difolco04},
although these are also consistent with those found by \cite{drakesmith93},
to within the combined uncertainties. Compared with the Sun, $\epsilon$~Eri
has a shorter rotational period (11.68~d) 
\citep{donahue96}, a larger spatially averaged magnetic 
field (165~G) \citep{rueedi97} and larger stellar surface emission 
line fluxes. It is, therefore, ideal for studies of a stellar outer atmosphere
under conditions of higher mean magnetic activity than the Sun. 

Observations with {\it IRAS} showed that $\epsilon$~Eri has an 
infrared excess, indicating the presence of dust; more recently a planet and
a debris disk have been detected \citep{hatzes00,greaves05}. Further studies of
the dust have been made by \cite{difolco07}.
Combining the above rotational period with $v \sin i = 1.7$~km~s$^{-1}$ and
$R_* = 0.743 R_{\odot}$ \citep{difolco04}, leads to $i = 32^\circ$, similar
to the value of 25$^\circ$ suggested by \cite{greaves05} for the inclination
of the debris disc to the plane of the sky. 

Here we make use of observations at ultraviolet to X-ray wavelengths, as 
summarized in Table~\ref{tab2}. In previous work we used
observations with the Space Telescope Imaging Spectrograph (STIS) on the 
{\it Hubble Space Telescope} to identify forbidden lines of Fe\,{\sc xii}
\citep{jordan01a} and to determine the electron pressure ($P_{\rm e}$) in the
transition region \citep{jordan01b}. Including observations with the 
{\it Far Ultraviolet Spectroscopic Explorer} ({\it FUSE}), we have modelled
 the chromosphere and transition region, up to an electron temperature 
($T_{\rm e}$) of $\simeq 3 \times 10^5$~K \citep{simjordan05}. 

Observations with the {\it Extreme Ultraviolet Explorer} ({\it EUVE}) have
been analysed by a number of authors. \cite{schmitt96} used lines of 
iron to derive the emission measure distribution (EMD) at above $T_{\rm e} 
\simeq 6 \times 10^5$~K and also placed limits on the electron density 
($n_{\rm e}$) in the upper transition region from lines of Fe\,{\sc xiii} and 
Fe\,{\sc xiv}. \cite{laming96} used the same spectra to derive
emission measures from lines of a range of elements and concluded that if
any trends in abundances with the first ionization potential (FIP) were
present, they were not significantly larger than in the solar corona. 
\cite{sf03} included further {\it EUVE} spectra and also derived the
EMD. They found much larger values of $n_{\rm e}$ from lines of Fe\,{\sc xix}
and Fe\,{\sc xxi} that are now superseded. \cite{simjordan03a}
used the line fluxes measured by \cite{schmitt96} to derive the EMD from
the iron lines alone, adopting a more recent (smaller) value of the
interstellar absorption by \cite{dring97} and using CHIANTI (v4)
\citep{dere97,young03} for line excitation atomic data. They compared the 
apparent EMD with that calculated from an energy balance in which the net 
thermal conductive flux was set equal to the local radiative energy loss and 
hence derived the fractional emitting area as a function of $T_{\rm e}$ 
($\simeq 0.2$ at $2 \times 10^5$~K).  

High-resolution X-ray observations of $\epsilon$~Eri have been made with both
 the Low Energy Transmission Grating Spectrograph (LETGS) on {\it Chandra} 
and the Reflection Grating Spectrometer (RGS) on {\it XMM-Newton}.
The former observations have been reduced and analysed by \cite{sf04} and 
\cite{wood06},
who derived the EMD and relative element abundances, making use of the
Astrophysical Plasma Emission Database (APED v1.3) \citep{smith01} and CHIANTI
(v4.2), respectively.
Because there have been changes in the atomic
data available, we have made our own analysis of the above spectra and the 
{\it EUVE} spectra, using CHIANTI (v5.2) \citep{landi06}. We have also
made our own measurements of the X-ray line fluxes.
$\epsilon$~Eri was included in a survey of stellar coronal densities
by \cite{ness02}, using LETGS spectra, and in a study of coronal
opacities by \cite{ness03a}, using LETGS and RGS spectra.
Comparisons with these earlier results are made in later sections.   

The first aim of the present work is to use all the available observations and
up-to-date atomic data to derive the mean EMD, the relative element abundances
and the electron densities and pressures. The mean EMD is then used in 
conjunction with new theoretical EMDs to refine the earlier values
of the mean emitting areas. 

Section~\ref{S2} describes the X-ray observations used and the data reduction.
Limits on $n_{\rm e}$ and $P_{\rm e}$ are discussed in Section~\ref{S3}, since these are
required in deciding on the identification of some lines and in the analysis 
of some line fluxes. The identifications of particular lines are discussed in 
Section~\ref{S4}, in conjunction with the emission measures derived and some 
issues related to the atomic data. Section~\ref{S5} describes how the mean
 EMD and the relative element abundances were derived. In Section~\ref{S6}, 
the theoretical EMD and the fractional areas occupied by the emitting regions
 are derived, using the method set out in \cite{simjordan03a}. Our main 
conclusions are summarized in Section~\ref{S7}.

% Table 1
\begin{table}
\caption{\label{tab1}Stellar parameters adopted.}
\begin{tabular}{p{.8cm}cccc}
\hline
Distance$^a$ & Mass$^b$       & Radius$^b$         & $\log g_{*}^b$  & $\log
N_{\rm H}^c$ \\  
  (pc)       & ($M_{\odot}$)  & ($R_{\odot}$)      &                 &       
 \\
\hline
3.218        & $0.9\,\pm\,0.1$ &  $0.743\,\pm\,0.01$ & $4.65\,\pm\,0.1$ & $17.88\,\pm\,0.07$\\
\hline
\end{tabular}

$^a$ The {\it Hipparcos} catalogue \cite{esa97}. \\
$^b$ \cite{difolco04}; $g_*$ in cm~s$^{-2}$. \cite{ap04} give a value of
$\log g_{*} = 4.621$.  \\
$^c$ \cite{dring97}; $N_{\rm H}$ in cm$^{-2}$.
\end{table} 

\section{Observations and data reduction}
\label{S2}
\subsection{X-ray observations}
\label{S2.1}
We use the {\it Chandra} LETGS spectra (ObsID 1869) that had an exposure 
time of 105.3~ks. We have also examined 
the spectra obtained with the {\it XMM-Newton} RGS instrument (ObsID 
0112880501) that have an exposure time of 13~ks. These spectra 
are available from the {\it Chandra} and {\it XMM-Newton} archives.  

The LETGS projects the dispersed spectrum onto a microchannel plate detector.
The detector is placed behind the grating in a manner such that the 
non-dispersed photons (zeroth order) are recorded in the middle of the 
detector, with the two dispersion directions appearing as spectra in the plus
and minus directions. There is overlap between the first-order spectra and 
the various higher-order spectra of shorter wavelengths and care must be taken
to exclude higher order lines or to account for line blending when this occurs
in important first-order lines. In the lists of lines given in
Tables~\ref{tab3} and \ref{tab4}, purely higher-order lines are excluded and 
blends with first-order lines are noted.

\subsection{Data reduction}
\label{S2.2}
We extracted the LETGS spectra on the plus and minus sides separately and
 calculated the effective areas using the standard CIAO tools (v3.2). The 
measurement of line fluxes was carried out with the CORA program developed by
\cite{newi02}. This accounts for the particular problems of low-count 
photon statistics \citep[see, e.g.,][]{ness01}. After trying several
approaches we used a fixed line width (FWHM- Full width at half maximum) of
0.060~\AA, since intrinsic stellar line widths are not resolvable. 

As noted by others \citep[see the extensive discussion in][]{chung04},
the apparent wavelengths of lines can differ between the plus and minus 
sides, leading to larger linewidths when the spectra in the plus and minus 
side spectra are summed. The line fluxes given in Tables~\ref{tab3} and
about half of those in 
\ref{tab4} are derived from the summed spectra, using variable line 
widths of between 0.053 and 0.080~\AA, although the larger widths usually 
occur only at wavelengths above 90~\AA. We have checked that these fluxes do 
not differ significantly from the averages of the fluxes in 
the individual plus and minus side spectra. Some lines are not observed, or
are poorly observed, in one of the two spectra. In this case only one spectrum
is used; the wavelengths of such lines are given
in italic script in Table~\ref{tab4}. We also give the effective areas 
adopted so that the original count rates can be recovered.
 
All the line fluxes have then been corrected for absorption in the 
interstellar medium (ISM) using 
a hydrogen column density of $\log N_{\rm H}$(cm$^{-2}$) = 17.88 from 
\cite{dring97}, lower than that used by \cite{sf03} who 
adopted a value of 18.1. The absorption model also includes He\,{\sc i}, 
He\,{\sc ii} and other abundant elements. Uncertainties in the line fluxes 
arise not only from the statistical measurement errors calculated by the 
CORA program, but also from the line widths adopted and the source continuum.
 The line widths in the LETGS spectra can be approximated by Moffat profiles 
\citep[Lorentzians with an exponent $\beta$ = 2.4;][]{drake04}. The source 
continuum can be accounted for in CORA by adding a background (in 
counts~s$^{-1}$~\AA$^{-1}$) to the line templates. By varying the source 
background we found that in many cases, the dominant source of the
uncertainty in line fluxes arises from the choice of the source background.
The specific case of the lines of O\,{\sc vii} is discussed in 
Section~\ref{S3}.

The APED \citep{smith01} and CHIANTI (v5.2) \citep{landi06} data bases
were consulted in making the line identifications (see also Section~\ref{S4}).
 Lines marked with {\it b} are possibly blended but the identification of 
the main 
additional contributor is not certain and these lines are not used in the 
analyses of line fluxes. Blending between first-order lines occurs in several 
 important cases. The methods used to find the relative contributions are 
discussed in Section~\ref{S4}. Fluxes were measured for all lines, but since 
our aim is to establish a reliable EMD a number of weak lines are not 
included in Tables~\ref{tab3} and \ref{tab4}, unless they have particular 
significance for our studies. A fuller list of lines present has been 
published by \cite{sf04}.

There are systematic differences between laboratory and observed wavelengths 
above about 80~\AA\ (where the observed wavelengths are too large), owing to 
the treatment of detector plate gaps \citep[see][]{chung04}. Since we are not
analyzing line shifts and are using only well identified lines in this region,
we have not attempted to correct these wavelengths. 

In reducing the RGS1 and RGS2 spectra we used Lorentzian line profiles to 
approximate the instrumental line profiles and used a fixed FWHM of 0.06~\AA. 
The only use we have made of these spectra is in comparisons between LETGS and
 RGS fluxes where line blending is not significant, because the instrumental 
wings to the lines prevents accurate deblending.   

\subsection{Other observations; potential variability of $\epsilon$~Eri}
\label{S2.3}
In later sections we will be using spectra obtained with the {\it EUVE},
STIS and {\it FUSE}. Table~\ref{tab2} gives the dates on which these
observations were made.

% Table 2
\begin{table}
\caption{\label{tab2}Dates of observations used.}
\begin{tabular}{lc}
\hline
Instrument & Date            \\
\hline
LETGS      & 2001 March 21    \\
RGS        & 2003 January 19  \\
{\it EUVE} & 1993 October 22/23 \\
STIS       & 2000 March  9      \\
{\it FUSE} & 2000 December 8     \\
\hline
\end{tabular}
\end{table}

\cite{baliunas83,baliunas95} have studied the Ca\,{\sc ii} 
emission lines 
over various time scales and find no clear single activity cycle. The monthly
 variations in the S-index cover a total amplitude of $\simeq 24$~per cent, 
so that the chromospheric emission does not show substantial variations. 
\cite{simjordan05} found no significant differences between the fluxes in 
transition region lines observed with the {\it International Ultraviolet 
Explorer} ({\it IUE}) in 1981 and with the STIS in 2000. 

%  Table 3
\begin{table}
\renewcommand{\arraystretch}{1.1}
\caption{\label{tab3}Line fluxes measured from the LETG spectrum of 
$\epsilon$\,Eri. See Section~\ref{S2} for details.
\vspace{-.1cm}}
\begin{flushleft}
{\scriptsize
\begin{tabular}{p{.3cm}p{1.cm}p{.4cm}crp{1.8cm}}
\hline
$\lambda$& Flux$^{a}$           &A$_{\rm eff}$&  $\lambda^{e}$ &  Ion   &
 Transition \\
(\AA) &                         &   (cm$^2$)  &      (\AA)    &        &
  \\
\hline
6.65  &\mbox{8.39\,$\pm$\,2.77} & 44.1   & 6.65    & Si\,{\sc xiii}  & 
 \mbox{\hspace{-.3cm}\tiny 1s$^2\,^1$S$_0$--1s2p\,$^1$P$_1$}\\
\multicolumn{2}{l}{\ --6.74}&            & 6.69    & Si\,{\sc xiii}  &  
\mbox{\hspace{-.3cm}\tiny 1s$^2\,^1$S$_{0}$--1s2p\,$^3$P$_{1,2}$}\\
&  &                                     & 6.74    & Si\,{\sc xiii}  &  
\mbox{\hspace{-.3cm}\tiny 1s$^2\,^1$S$_{0}$--1s2s\,$^3$S$_{1}$}\\
8.42  &\mbox{3.68\,$\pm$\,0.66} & 38.2   & 8.42    & Mg\,{\sc xii}   &  
\mbox{\hspace{-.3cm}\tiny 1s\,$^2$S$_{1/2}$--2p\,$^2$P$_{1/2,3/2}$}\\
9.17  &\mbox{12.6\,$\pm$\,2.5}  & 32.5   & 9.17    & Mg\,{\sc xi}    &  
\mbox{\hspace{-.3cm}\tiny 1s$^2\,^1$S$_0$--1s2p\,$^1$P$_1$}\\
\multicolumn{2}{l}{\ --9.31}&            & 9.23    & Mg\,{\sc xi}    &  
\mbox{\hspace{-.3cm}\tiny 1s$^2\,^1$S$_{0}$--1s2p\,$^3$P$_{1,2}$}\\
&  &                                     & 9.31    & Mg\,{\sc xi}    &
\mbox{\hspace{-.3cm}\tiny 1s$^2\,^1$S$_{0}$--1s2s\,$^3$S$_{1}$}\\
10.23 &\mbox{4.20\,$\pm$\,0.75} & 28.7  & 10.24    & Ne\,{\sc x}     & 
\mbox{\hspace{-.3cm}\tiny 1s\,$^2$S$_{1/2}$--3p\,$^2$P$_{1/2,3/2}$}\\
11.27 &\mbox{4.69\,$\pm$\,0.74} & 28.5  & 11.25    & Fe\,{\sc xvii} & 
\mbox{\hspace{-.3cm}\tiny 2p$^6\,^1$S$_{0}$--2p$^5$($^2$P)5d\,$^3$D$_{1}~^b$}\\
11.55 &\mbox{5.58\,$\pm$\,0.75} & 29.1  & 11.55    & Ne\,{\sc ix} &
\mbox{\hspace{-.3cm}\tiny 2s$^2\,^1$S$_{0}$--1s3p\,$^1$P$_{1}~^b$}\\
12.14 &\mbox{29.5\,$\pm$\,1.4}  & 28.7  & 12.14    & Ne\,{\sc x}     & 
\mbox{\hspace{-.3cm}\tiny 1s\,$^2$S$_{1/2}$--2p\,$^2$P$_{1/2,3/2}$}\\
      & $\sim 27$\%$^c$         &       & 12.12    & Fe\,{\sc xvii}  & 
\mbox{\hspace{-.3cm}\tiny 2p$^6\,^1$S$_0$--2p$^5$4d\,$^1$P$_1$}\\
12.29 &\mbox{7.18\,$\pm$\,0.80} & 28.5  & 12.26    & Fe\,{\sc xvii}  &  
\mbox{\hspace{-.3cm}\tiny 2p$^6\,^1$S$_{0}$--2p$^5$($^2$P)4d\,$^3$D$_{1}$}\\
13.45 &\mbox{22.9\,$\pm$\,1.05}$^d$ & 29.4  & 13.45    & Ne\,{\sc ix}   &  
\mbox{\hspace{-.3cm}\tiny 1s$^2\,^1$S$_0$--1s2p\,$^1$P$_1$}\\
13.55 &\mbox{7.57\,$\pm$\,0.60}$^d$ & 29.4  & 13.55    & Ne\,{\sc ix}  &  
\mbox{\hspace{-.3cm}\tiny 1s$^2\,^1$S$_0$--1s2p\,$^3$P$_{1,2}$}  \\
13.70 &\mbox{16.1\,$\pm$\,0.9}$^d$  & 29.4  & 13.70    & Ne\,{\sc ix}   &  
\mbox{\hspace{-.3cm}\tiny 1s$^2\,^1$S$_0$--1s2s\,$^3$S$_{1}$}\\
13.83 &\mbox{3.67\,$\pm$\,0.41} & 29.4  & 13.82    & Fe\,{\sc xvii}  & 
\mbox{\hspace{-.3cm}\tiny 2s$^2$2p$^6\,^1$S$_0$--2s2p$^6$3p$^1$P$_1$ $^b$} \\
14.21 &\mbox{10.9\,$\pm$\,0.9}  & 29.5  & 14.21    & Fe\,{\sc xviii}  & 
\mbox{\hspace{-.3cm}\tiny 2p$^5\,^2$P$_{3/2}$--2p$^4$($^1$D)3d\,$^2$D$_{5/2}$}\\
\multicolumn{2}{l}{\ --14.26 }  &       & 14.26    & Fe\,{\sc xviii} & 
\mbox{\hspace{-.3cm}\tiny 2p$^5\,^2$P$_{3/2}$--2p$^4$3d$^{2}$S$_{1/2}$ $^b$} \\
14.38 &\mbox{4.19\,$\pm$\,0.69} & 29.6  & 14.37    & Fe\,{\sc xviii} &  
\mbox{\hspace{-.3cm}\tiny 2p$^5\,^2$P$_{3/2}$--2p$^4$($^3$P)3d\,$^2$D$_{5/2}$
$^b$}\\
15.02 &\mbox{52.1\,$\pm$\,1.6 } & 30.3  & 15.02    & Fe\,{\sc xvii}  &  
\mbox{\hspace{-.3cm}\tiny 2p$^6\,^1$S$_{0}$--2p$^5$($^2$P)3d\,$^1$P$_{1}$}\\
15.19 &\mbox{7.98\,$\pm$\,0.84} & 30.4  & 15.18    & O\,{\sc viii}   & 
\mbox{\hspace{-.3cm}\tiny 1s$\,^2$S$_{1/2}$--4p$\,^2$P$_{1/2,3/2}$ $^b$}\\
15.27 &\mbox{21.8\,$\pm$\,1.1}  & 30.5  & 15.26    & Fe\,{\sc xvii}  & 
\mbox{\hspace{-.3cm}\tiny 2p$^6\,^1$S$_{0}$--2p$^5$($^2$P)3d\,$^3$D$_{1}$$^b$}\\
16.01 &\mbox{17.1\,$\pm$\,1.0}  & 29.9  & 16.01    & O\,{\sc viii}   &  
\mbox{\hspace{-.3cm}\tiny 1s\,$^2$S$_{1/2}$--3p\,$^2$P$_{1/2,3/2}$}\\
      &$\sim 19$\%$^c$          &       & 16.01    & Fe\,{\sc xviii} & 
\mbox{\hspace{-.3cm}\tiny 2p$^5\,^2$P$_{3/2}$--2p$^4$($^3$P)3s\,$^2$P$
_{3/2}$}\\
16.09 &\mbox{5.80\,$\pm$\,0.67} & 30.2  & 16.08    & Fe\,{\sc xviii} &  
\mbox{\hspace{-.3cm}\tiny 2p$^5\,^2$P$_{3/2}$--2p$^4$($^3$P)3s\,$^4$P$_{5/2}$}\\
16.77 &\mbox{30.4\,$\pm$\,1.2}  & 30.6  & 16.78    & Fe\,{\sc xvii}  & 
\mbox{\hspace{-.3cm}\tiny 2p$^6\,^1$S$_{0}$--2p$^5$($^2$P)3s\,$^1$P$_{1}$}\\
17.05 &\mbox{77.0\,$\pm$\,3.0}  & 25.7  & 17.05    & Fe\,{\sc xvii}  & 
\mbox{\hspace{-.3cm}\tiny 2p$^6\,^1$S$_{0}$--2p$^5$($^2$P)3s\,$^3$P$_{1}$}\\
\multicolumn{2}{l}{\ --17.10 }  &       & 17.10    & Fe\,{\sc xvii}  & 
\mbox{\hspace{-.3cm}\tiny 2p$^6\,^1$S$_{0}$--2p$^5$($^2$P)3s\,$^3$P$_{2}$}\\
18.63 &\mbox{7.79\,$\pm$\,1.26} & 26.3  & 18.63    & O\,{\sc vii}    &  
\mbox{\hspace{-.3cm}\tiny 1s$^2\,^1$S$_{0}$--1s3p\,$^1$P$_{1}$ $^b$}\\
18.97 &\mbox{88.2\,$\pm$\,1.9}  & 26.5  & 18.97    & O\,{\sc viii}   & 
\mbox{\hspace{-.3cm}\tiny 1s\,$^2$S$_{1/2}$--2p\,$^2$P$_{1/2,3/2}$}\\
21.61 &\mbox{41.5\,$\pm$\,1.5}  & 17.3  & 21.60    & O\,{\sc vii}    &  
\mbox{\hspace{-.3cm}\tiny 1s$^2\,^1$S$_0$--1s2p\,$^1$P$_1$}\\
21.81 &\mbox{9.60\,$\pm$\,0.83} & 17.0  & 21.81    & O\,{\sc vii}    &  
\mbox{\hspace{-.3cm}\tiny 1s$^2\,^1$S$_{0}$--1s2p\,$^3$P$_{1,2}$}\\
22.11 &\mbox{27.3\,$\pm$\,1.3}  & 17.0  & 22.10    & O\,{\sc vii}    &  
\mbox{\hspace{-.3cm}\tiny 1s$^2\,^1$S$_{0}$--1s2s\,$^3$S$_{1}$}\\
24.79 &\mbox{10.8\,$\pm$\,0.9}  & 16.8  & 24.78    & N\,{\sc vii}    &  
\mbox{\hspace{-.3cm}\tiny 1s\,$^2$S$_{1/2}$--2p\,$^2$P$_{1/2,3/2}$}\\
28.47 &\mbox{3.53\,$\pm$\,0.55} & 15.6  & 28.47    & C\,{\sc vi}     &  
\mbox{\hspace{-.3cm}\tiny 1s$^2$S$_{1/2}$--3p\,$^2$P$_{1/2,3/2}$}\\
28.79 &\mbox{3.56\,$\pm$\,0.55} & 15.3  & 28.79    & N\,{\sc vi}     &  
\mbox{\hspace{-.3cm}\tiny 1s$^2\,^1$S$_0$--1s2p\,$^1$P$_1$}\\
29.09 &\mbox{$<2.50$}           & 15.0  & 29.08    & N\,{\sc vi}     &  
\mbox{\hspace{-.3cm}\tiny 1s$^2\,^1$S$_{0}$--1s2p\,$^3$P$_{1,2}$}\\
29.54 &\mbox{$<2.80$}           & 14.1  & 29.53    & N\,{\sc vi}     &  
\mbox{\hspace{-.3cm}\tiny 1s$^2\,^1$S$_{0}$--1s2s\,$^3$S$_{1}$}\\
30.45 &\mbox{4.37\,$\pm$\,0.64} & 12.1  & 30.47    & S\,{\sc xiv}    &  
\mbox{\hspace{-.3cm}\tiny 2s\,$^2$S$_{1/2}$--3p\,$^2$P$_{1/2,3/2}$}\\
&\mbox{}                        &       & 30.45    & Ca\,{\sc xi}    &  
\mbox{\hspace{-.3cm}\tiny 2p$^6\,^1$S$_0$--2p$^5$3d\,$^1$P$_1$}\\
32.24 &\mbox{2.47\,$\pm$\,0.52} & 13.5  & 32.24    & S\,{\sc xiii}  &  
\mbox{\hspace{-.3cm}\tiny 2s$^2$\,$^1$S$_0$--2s3p\,$^1$P$_1$}   $^b$ \\
32.56 &\mbox{3.70\,$\pm$\,0.55} & 13.3  & 32.56    & S\,{\sc xiv}    &  
\mbox{\hspace{-.3cm}\tiny 2p\,$^2$P$_{3/2}$--3d\,$^2$D$_{3/2,5/2}$}\\
33.55 &\mbox{2.47\,$\pm$\,0.49} & 12.8  & 33.55    & S\,{\sc xiv}    &  
\mbox{\hspace{-.3cm}\tiny 2p\,$^2$P$_{3/2}$--3s\,$^2$S$_{1/2}$}\\
33.74 &\mbox{17.1\,$\pm$\,1.0}  & 12.8  & 33.74    & C\,{\sc vi}     &  
\mbox{\hspace{-.3cm}\tiny 1s\,$^2$S$_{1/2}$--2p\,$^2$P$_{1/2,3/2}$}\\
35.69 &\mbox{4.17\,$\pm$\,0.55} & 12.3  & 35.67    & S\,{\sc xiii}   &  
\mbox{\hspace{-.3cm}\tiny 2s2p\,$^1$P$_1$--2s3d\,$^1$D$_2$ $^b$} \\
37.61 &\mbox{2.14\,$\pm$\,0.49} & 10.4  & 37.60    & S\,{\sc xiii}   &  
\mbox{\hspace{-.3cm}\tiny 2s2p\,$^1$P$_1$--2s3s\,$^1$S$_0$ $^b$} \\
40.27 &\mbox{3.35\,$\pm$\,0.78$^d$} &  5.5  & 40.27    & C\,{\sc v}      & 
\mbox{\hspace{-.3cm}\tiny 1s$^2\,^1$S$_0$--1s2p\,$^1$P$_1$}\\
43.76 &\mbox{2.89\,$\pm$\,0.27} & 25.9  & 43.76    & Si\,{\sc xi}    & 
\mbox{\hspace{-.3cm}\tiny 2s$^2\,^1$S$_0$--2s3p\,$^1$P$_1$ $^b$} \\
44.02 &\mbox{3.75\,$\pm$\,0.29} & 26.1  & 44.02    & Si\,{\sc xii}   &  
\mbox{\hspace{-.3cm}\tiny 2p\,$^2$P$_{1/2}$--3d\,$^2$D$_{3/2}$ $^b$} \\
44.17 &\mbox{5.47\,$\pm$\,0.34} & 26.1  & 44.18    & Si\,{\sc xii}   &  
\mbox{\hspace{-.3cm}\tiny 2p\,$^2$P$_{3/2}$--3d\,$^2$D$_{3/2,5/2}$}\\
45.51 &\mbox{1.30\,$\pm$\,0.21} & 25.7  & 45.52    & Si\,{\sc xii}   &  
\mbox{\hspace{-.3cm}\tiny 2p\,$^2$P$_{1/2}$--3s\,$^2$S$_{1/2}$}\\
45.69 &\mbox{2.01\,$\pm$\,0.24} & 25.6  & 45.69    & Si\,{\sc xii}   &  
\mbox{\hspace{-.3cm}\tiny 2p\,$^2$P$_{3/2}$--3s\,$^2$S$_{1/2}$}\\
\hline
\end{tabular}
\\ 
$^a$ Fluxes are in $10^{-14}$~erg~cm$^{-2}$~s$^{-1}$. \\
$^b$ Likely to be blended. \\
$^c$ The percentage contribution to the total flux given. \\
$^d$ Fluxes after deblending. \\
$^e$ Wavelengths from CHIANTI (v5.2).\\
}
\renewcommand{\arraystretch}{1}
\end{flushleft}
\end{table}

%  Table 4
\begin{table}
\renewcommand{\arraystretch}{1.1}
\caption{\label{tab4}Continuation from Table~\ref{tab3}.} 
\begin{flushleft}
{\scriptsize
\begin{tabular}{p{.5cm}p{1.cm}p{.4cm}p{.5cm}rp{1.8cm}}
\hline
$\lambda$ & Flux$^{a}$ & A$_{\rm eff}$ & $\lambda^{e}$ & Ion & Transition   \\
(\AA) &                          &(cm$^2$)~& (\AA)  &                 &   \\
\hline
49.21       &\mbox{3.79\,$\pm$\,0.29}   & 24.4 &  49.22 & Si\,{\sc xi}  &  
\mbox{\hspace{-.3cm}\tiny 2s2p\,$^1$P$_1$--2s3d\,$^1$D$_2$}\\
{\it 50.36}$^d$&\mbox{4.96\,$\pm$\,0.49}& 10.9 &  50.36 & Fe\,{\sc xvi} &  
\mbox{\hspace{-.3cm}\tiny 3s\,$^2$S$_{1/2}$--4p\,$^2$P$_{3/2}$}\\
       &$\sim 29$\%$^c$           &       &  50.34  & 3$^{\rm rd}$ & order 
16.78\,\AA \\
{\it 50.55} &\mbox{3.75\,$\pm$\,0.44}  &  10.9 &  50.52 &  Si\,{\sc x} & 
\mbox{\hspace{-.3cm}\tiny 2p\,$^2$P$_{1/2}$--3d\,$^2$D$_{3/2}$}\\
            &$\sim 51$\%$^c$           &       &  50.57 & Fe\,{\sc xvi} &  
\mbox{\hspace{-.3cm}\tiny 3s\,$^2$S$_{1/2}$--4p\,$^2$P$_{1/2}$}\\
{\it 50.69} &\mbox{2.83\,$\pm$\,0.39}  &  10.9 &  50.69 &   Si\,{\sc x}  &  
\mbox{\hspace{-.3cm}\tiny 2p\,$^2$P$_{3/2}$--3d\,$^2$D$_{5/2,3/2}$}\\
{\it 52.32} &\mbox{2.16\,$\pm$\,0.36}  &  10.4 &  52.30  & Si\,{\sc xi}  & 
 \mbox{\hspace{-.3cm}\tiny 2s2p\,$^1$P$_1$--2s3s\,$^1$S$_0$}\\
{\it 52.90} &\mbox{2.24\,$\pm$\,0.37}  &  10.4 & 52.91   &  Fe\,{\sc xv} &
  \mbox{\hspace{-.3cm}\tiny 3s$^2\,^1$S$_0$--3s4p\,$^1$P$_1$}\\
{\it 54.14} & \mbox{2.84\,$\pm$\,0.38} &  10.2 & 54.13 & Fe\,{\sc xvi} &
  \mbox{\hspace{-.3cm}\tiny 3p\,$^2$P$_{1/2}$--4d\,$^2$D$_{3/2}$}    \\
{\it 54.72} & \mbox{4.78\,$\pm$\,0.46} &  10.1 & 54.71 & Fe\,{\sc xvi} &
 \mbox{\hspace{-.3cm}\tiny 3p\,$^2$P$_{3/2}$--4d\,$^2$D$_{5/2,3/2}$}  \\
{\it 57.90} &\mbox{2.11\,$\pm$\,0.48}  &  9.2  &  57.92  &    Mg\,{\sc x}  & 
 \mbox{\hspace{-.3cm}\tiny 2s\,$^2$S$_{1/2}$--3p\,$^2$P$_{1/2,3/2}$}\\
{\it 59.40} &\mbox{2.97\,$\pm$\,0.44}  &  8.0  &  59.40  &   Fe\,{\sc xv}  &  
\mbox{\hspace{-.3cm}\tiny 3s3p\,$^1$P$_1$--3s4d\,$^1$D$_2$}\\
{\it 62.88} &\mbox{2.73\,$\pm$\,0.43}  &  7.7  &  62.87  &    Fe\,{\sc xvi} &
  \mbox{\hspace{-.3cm}\tiny 3p\,$^2$P$_{1/2}$--4s\,$^2$S$_{1/2}$}\\
 {\it 63.15} &\mbox{1.07\,$\pm$\,0.34} &  7.7  &  63.15  &    Mg\,{\sc x}   &
  \mbox{\hspace{-.3cm}\tiny 2p$\,^2$P$_{1/2}$--3d\,$^2$D$_{3/2}$}\\
 {\it 63.31} &\mbox{2.80\,$\pm$\,0.44} &  7.7  &  63.31  &    Mg\,{\sc x}  & 
 \mbox{\hspace{-.3cm}\tiny 2p\,$^2$P$_{3/2}$--3d\,$^2$D$_{5/2,3/2}$}\\
 {\it 63.73} &\mbox{6.20\,$\pm$\,0.57} &  7.6  &  63.71  &    Fe\,{\sc xvi} &
  \mbox{\hspace{-.3cm}\tiny 3p\,$^2$P$_{3/2}$--4s\,$^2$S$_{1/2}$}\\
 {\it 65.86} &\mbox{1.06\,$\pm$\,0.37} &  7.3  &  65.85  &    Mg\,{\sc x} &
 \mbox{\hspace{-.3cm}\tiny 2p\,$^2$P$_{3/2}$--3s\,$^2$S$_{1/2}$}\\
 {\it 66.25} &\mbox{3.78\,$\pm$\,0.51} &  7.2  &  66.25  &   Fe\,{\sc xvi} &
  \mbox{\hspace{-.3cm}\tiny 3d\,$^2$D$_{3/2}$--4f\,$^2$F$_{5/2}$} \\
             &                         &       &  66.30  & 3$^{\rm rd}$ &
  order 22.10~\AA       \\
 {\it 66.35} &\mbox{5.91\,$\pm$\,0.60} &  7.1  &  66.36  &    Fe\,{\sc xvi} &
 \mbox{\hspace{-.3cm}\tiny 3d\,$^2$D$_{5/2}$--4f\,$^2$F$_{7/2,5/2}$} \\
  69.60 &\mbox{1.07\,$\pm$\,0.25}      & 14.3  &         &   & $b$\\  
  69.68 &\mbox{5.47\,$\pm$\,0.41}      & 14.3  &  69.68  &   Fe\,{\sc xv} & 
\mbox{\hspace{-.3cm}\tiny 3s3p\,$^1$P$_1$--3s4s\,$^1$S$_0$}\\
  72.32 &\mbox{$<1.14$}                & 13.5  &  72.31  &  Mg\,{\sc ix} & 
\mbox{\hspace{-.3cm}\tiny 2s2p\,$^1$P$_1$--2s3d\,$^1$D$_2$}\\
  73.48 &\mbox{2.70\,$\pm$\,0.53}      & 13.0  &  73.47  &  Fe\,{\sc xv}  &
  \mbox{\hspace{-.3cm}\tiny 3s3d\,$^1$D$_2$--3s4f\,$^1$F$_3$} \\
        &                              &       & 73.48   & Ne\,{\sc viii} & 
  \mbox{\hspace{-.3cm}\tiny 2p\,$^2$P$_{1/2}$--4d\,$^2$D$_{3/2}$}\\
  76.04 &\mbox{1.32\,$\pm$\,0.27}      & 12.2  & 76.02   & Fe\,{\sc xiv}  & 
 \mbox{\hspace{-.3cm}\tiny 3d\,$^2$D$_{3/2}$--4f\,$^2$F$_{5/2,3/2}$}  \\
  76.13 &\mbox{1.21\,$\pm$\,0.27}      & 12.2  & 76.15   & Fe\,{\sc xiv}  &
 \mbox{\hspace{-.3cm}\tiny 3d\,$^2$D$_{5/2}$--4f\,$^2$F$_{7/2}$}   \\
  76.53 &\mbox{1.61\,$\pm$\,0.28}      & 12.0  &         & ? & \\ 
        &                              &       & 76.50   & Fe\,{\sc xvi}  &
 \mbox{\hspace{-.3cm}\tiny 3d\,$^2$D$_{5/2}$--4p\,$^2$P$_{3/2}~^b$} \\
 77.74 &\mbox{$<0.63$}                 & 11.7  & 77.74   &  Mg\,{\sc ix} & 
\mbox{\hspace{-.3cm}\tiny 2s2p\,$^1$P$_1$--2s3s\,$^1$S$_0$}\\
 88.11 &\mbox{5.07\,$\pm$\,0.89}       &  9.5  & 88.08   & Ne\,{\sc viii}  &
  \mbox{\hspace{-.3cm}\tiny 2s\,$^2$S$_{1/2}$--3p\,$^2$P$_{3/2}$}\\
 \multicolumn{2}{l}{\ --88.15}         &       & 88.12   & Ne\,{\sc viii}  &
  \mbox{\hspace{-.3cm}\tiny 2s\,$^2$S$_{1/2}$--3p\,$^2$P$_{1/2}$}\\  
  93.97 &\mbox{7.08\,$\pm$\,0.51}      &  8.8  & 93.92   & Fe\,{\sc xviii} &
  \mbox{\hspace{-.3cm}\tiny 2s$^2$2p$^5\,^2$P$_{3/2}$--2s2p$^6\,^2$S$_{1/2}$}\\
  98.15 &\mbox{1.68\,$\pm$\,0.34}      &  7.6  & 98.12   & Ne\,{\sc viii}  &
  \mbox{\hspace{-.3cm}\tiny 2p\,$^2$P$_{1/2}$--3d\,$^2$D$_{3/2}$}\\
  98.28 &\mbox{3.90\,$\pm$\,0.45}      &  7.6  & 98.26   & Ne\,{\sc viii}  &
  \mbox{\hspace{-.3cm}\tiny 2p\,$^2$P$_{3/2}$--3d\,$^2$D$_{5/2}$}\\
&&&98.27 &       Ne\,{\sc viii} & \mbox{\hspace{-.3cm}\tiny 
2p\,$^2$P$_{3/2}$--3d\,$^2$D$_{3/2}$}\\
 101.62 &\mbox{1.43\,$\pm$\,0.36}  &  7.1  &  101.55 &  Fe\,{\sc xix}  &  
\mbox{\hspace{-.3cm}\tiny 2s$^2$2p$^4\,^3$P$_{2}$--2s2p$^5\,^3$P$_{1}$}\\
 104.00 &\mbox{2.57\,$\pm$\,0.41}  &  6.9  &  103.94 &  Fe\,{\sc xviii}  & 
 \mbox{\hspace{-.3cm}\tiny 2s$^2$2p$^5\,^2$P$_{1/2}$--2s2p$^6\,^2$S$_{1/2}$}\\
 108.42 &\mbox{2.54\,$\pm$\,0.41}  &  6.8  &  108.36 &  Fe\,{\sc xix} & 
 \mbox{\hspace{-.3cm}\tiny 2s$^2$2p$^4\,^3$P$_{2}$--2s2p$^5\,^3$P$_{2}$}\\
 132.93 &\mbox{$<4$}               &  3.9  &  132.84 &  Fe\,{\sc xx}  &  
\mbox{\hspace{-.3cm}\tiny 2s$^2$2p$^3\,^4$S$_{3/2}$--2s2p$^4\,^4$P$_{5/2}$}\\
 141.09 &\mbox{2.57\,$\pm$\,0.57}  &  4.0  &  141.04 &  Ca\,{\sc xii}  &  
\mbox{\hspace{-.3cm}\tiny 2s$^2$2p$^5\,^2$P$_{3/2}$--2s2p$^6\,^2$S$_{1/2}$}\\
 150.16 &\mbox{$<2.17$}            &  3.6  &  150.12 &  O\,{\sc vi}   &  
\mbox{\hspace{-.3cm}\tiny 2s\,$^2$S$_{1/2}$--3p\,$^2$P$_{3/2,1/2}$}\\
 152.25 &\mbox{2.82\,$\pm$\,0.57}  &  3.6  &  152.15 &  Ni\,{\sc xii}  &  
\mbox{\hspace{-.3cm}\tiny 3p$^5\,^2$P$_{3/2}$--3p$^4$($^3$P)3d\,
$^2$D$_{5/2}$}\\
        &                          &       &         &  Ni\,{\sc xii}  &
\mbox{\hspace{-.3cm}\tiny 3p$^5\,^2$P$_{3/2}$--3p$^4$($^3$P)3d\,
$^2$P$_{1/2}$}\\
 154.25 &\mbox{1.57\,$\pm$\,0.50}  &  3.6  &  154.16 &  Ni\,{\sc xii}  &
 \mbox{\hspace{-.3cm}\tiny 3p$^5\,^2$P$_{3/2}$--3p$^4$($^3$P)3d\,
 $^2$D$_{3/2}$}\\
{\it 171.17} &\mbox{13.0\,$\pm$\,1.7}&1.2  &  171.07 &  Fe\,{\sc ix} &
  \mbox{\hspace{-.3cm}{\tiny 3p$^6\,^1$S$_0$--3p$^5$3d\,$^1$P$_1$}}\\
\hline
\end{tabular}
\\
$^a$ Fluxes are in $10^{-14}$~erg~cm$^{-2}$~s$^{-1}$. \\
$^b$ Likely to be blended. \\
$^c$ The percentage contribution to the total flux given. \\
$^d$ Wavelengths in italics indicate flux measurements on one side only. \\
$^e$ Wavelengths from CHIANTI (v5.2).\\}
\renewcommand{\arraystretch}{1}
\end{flushleft}
\end{table}

Regarding variations in emission from the upper transition region and corona, 
\cite{sf03}, using the {\it EUVE} Deep Survey 
Imager, observed similar mean levels of counts in the 80~\AA\ to 180~\AA\ 
passband in 1993 and 1995,
although some variations of the order of $\pm 30$~per cent were observed over
 a time interval of hours. Similarly, \cite{sf04} show light curves obtained
from the LETGS observations that exhibit variations amounting
to about $\pm 21$~per cent over 28 hours. 
The only systematic difference between the LETGS
and {\it EUVE} observations is that the fluxes of lines of iron in stages 
of ionization greater than Fe\,{\sc xviii} are lower in the LETGS spectra,
suggesting that the star had less hot active region material at that time.    
For the lines observed in both the RGS1 and RGS2 spectra, the mean RGS 
fluxes are on average a factor of only 1.06 larger than those derived from
the LETGS spectra.  
We conclude that significant uncertainties should not be introduced by 
combining observations made on different dates.  

\section{The electron pressure}
\label{S3}

\cite{jordan01b} measured a number of density-sensitive line flux ratios
in their STIS spectrum of $\epsilon$~Eri and used these to investigate $P_{\rm e}$
and to assess the available atomic data (in CHIANTI v3.01 and relevant 
papers). Throughout the present work we define $P_{\rm e} = 
N_{\rm e} T_{\rm e}$~cm$^{-3}$~K.
 Values of $\log P_{\rm e} = 15.69\,\pm\,0.1$ at 
$\log T_{\rm e} \simeq 4.5$ and $\simeq 15.67$ at $\log T_{\rm e} 
\simeq 5.15$ were obtained  
from the transition region lines of Si\,{\sc iii} and O\,{\sc iv}, 
respectively, although for O\,{\sc iv}, the four independent ratios gave 
values of $\log P_{\rm e}$ between $15.28\,\pm\,0.08$ and $16.16\,\pm\,0.17$. The 
overall mean value of $\log P_{\rm e} = 15.68$ was 
consistent with the lower and upper limits provided by lines of C\,{\sc iii}, 
O\,{\sc v} and Fe\,{\sc xii}.

We have re-examined all the pressures derived using CHIANTI (v5.2), since 
some changes were made following the discussions in \cite{jordan01b}.
There have been no changes to the data for the lines of C\,{\sc iii} and 
Si\,{\sc iii}. The pressures found from the ratios involving the O\,{\sc iv}
1401-\AA\ line are now slightly lower ($\log P_{\rm e} = 16.01\,\pm\,0.17$ 
and $15.98\,\pm\,0.20$), but the discordant low pressures derived from ratios involving
the blended line at 1404~\AA\ are unchanged, and there are still problems
with the lines of S\,{\sc iv}. A pressure can now be found from the lines of
O\,{\sc v} at 1218~\AA\ and 1371~\AA, and is $\log P_{\rm e} = 
15.95^{+0.18}_{-0.24}$. If the pressures from the ratios involving the 
O\,{\sc iv} 1401-\AA\ line
are preferred, then the mean electron pressure at around $\log T_{\rm e} = 
5.3$ becomes $\simeq 15.97\,\pm\,0.2$. It should be noted that if material
at pressures up to about $\log P_{\rm e} = 16.30$ were present,
this could be detected from the line flux ratios used.

In models of the chromosphere and transition region by \cite{simjordan05}, a 
turbulent pressure term derived from the observed non-thermal line widths 
\citep{jordan01b}
is included in the equation of hydrostatic equilibrium. This results in small
 {\it increases} in $P_{\rm e}$ with increasing $T_{\rm e}$ above $3 \times 10^4$~K, 
where the Si\,{\sc iii} lines are formed. The pressures now derived from the
 transition region lines are consistent with this behaviour, to within the 
uncertainties given above. In our theoretical models of the upper transition 
region and inner corona \citep[and in those by][]{simjordan03a}, $P_{\rm e}$ 
continues to rise by a few per cent until $\log T_{\rm e} \simeq 5.8$ (see 
Section~\ref{S6}). Only a small difference in $P_{\rm e}$ is expected between 
$\log T_{\rm e} = 5.3$ and 6.3, where the lines of O\,{\sc vii} are mainly formed.

%Fig. 1
\begin{figure}
\resizebox{\hsize}{!}{\includegraphics{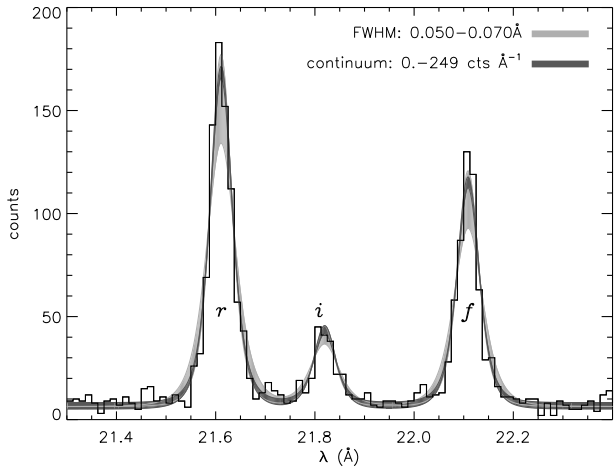}}
\caption{\label{f1}The spectral region including the O\,{\sc vii} triplet.
The light grey shade indicates the best fits when the
background is fixed at 10 counts \AA$^{-1}$ and the line width is varied
between values of 0.050~\AA\ -- 0.070~\AA. The dark grey shade indicates the
 best fits when the line width is fixed at 0.053~\AA\ and the background is 
varied between values of 0 -- 249 counts \AA$^{-1}$.}
\end{figure}

In the X-ray spectra, the ratio of the fluxes in the forbidden line (f) 
(1s2s~$^3$S -- 1s$^2$~$^1$S) and the intersystem line (i) (1s2p~$^3$P$_{1,2}$ 
-- 1s$^2$~$^1$S) in the He\,{\sc i}-like ions is potentially sensitive to
 $n_{\rm e}$ \citep{gj69}. Of the He\,{\sc i}-like ions observed with the LETGS,
 only O\,{\sc vii} has lines that are sufficiently strong and unblended to use
 in a measurement of $n_{\rm e}$. Because of the importance of the measured flux
 ratio we have varied the continuum level and line widths used in extracting
 the fluxes, in order to obtain realistic error bars. The spectrum and these
 fits are shown in Fig.~\ref{f1}. The mean observed ratio is 
2.88 $^{+0.57}_{-0.34}$. Provided the same widths are used for both lines,
 uncertainties in the widths have less effect than those in the choice of 
continuum. The range of the observed ratio is 2.535 -- 3.455, compatible with
 the ratio of 3.06 measured by \cite{sf04}, who allowed explicitly for 
contributions from what they regarded as weak unidentified lines. 

Using CHIANTI (v4.2), at $T_{\rm e} = 2 \times 10^6$~K, the observed ratio of 2.88 
leads to $\log P_{\rm e} = 16.35$, with a range of 16.54 -- 16.00. 
%Only values close to the lower limit are
The lower limit is consistent with the pressure found 
from the transition region lines at $\log T_{\rm e} = 5.3$. This pressure is 
essentially the same as the value of $\log P_{\rm e} = 16.33$ found in an
earlier analysis by \cite{ness02}, which was adopted by \cite{sf04}. 
 (Using atomic data from APED 
gives essentially the same results.) CHIANTI (v4.2) predicts a ratio of 3.9 
in the low-density limit, but does not include recombination to the n = 2 
levels, either directly or via cascades. Early work by \cite{gj73}
showed that including both radiative and di-electronic recombination tends to
increase the predicted ratio in the low-density limit, but by only a 
relatively small amount; collisional excitation followed by cascades is more
important. They predicted a low-density limit of 3.64. \cite{blum72} found 
larger effects from recombination, but according to \cite{gj73} they
 overestimated the contributions from di-electronic recombination.

The most recent version of CHIANTI (v5.2) gives the option of including
 radiative recombination as a process populating the excited states. 
The implementation of radiative recombination in CHIANTI (v5.2) is not 
correct, since it does not take into account recombination to the 
1s2p~$^3$P$_0$ level \citep{landi06} and hence omits an important 
process for populating the 1s2s~$^3$S level. This leads to lower values of the
f/i ratio at a given value of $n_{\rm e}$. In particular, the value of f/i at low
densities ($R_o$) becomes $\simeq 3.35$.

\cite{pordu00} have given radiative and di-electronic recombination 
rate coefficients and effective collision strengths for populating 
the n = 2 levels, including the effects of cascades from n $>$ 2 in all 
cases.  At $\log T_{\rm e} = 6.3$ their calculations predict a value of $R_o = 3.82$,
a little smaller than that found from CHIANTI (v4.2). Compared with the work 
by \cite{gj73}, \cite{pordu00} find larger contributions from collisional
 excitations, followed by cascades, to levels with n $>$ 2. Using the 
calculations by \cite{pordu00} (but neglecting the small contribution from 
recombination, since this causes only a small increase in $R_o$) leads to 
pressures that are similar to those from CHIANTI (v4.2) ($\log P_{\rm e} = 16.38$, 
with a range from 16.56 to 15.93). Here the lower limit is compatible with the
 pressure found at around $\log T_{\rm e} = 5.3$, without considering the error 
bars.  

At present, the origin of the higher optimum pressure found 
from the O\,{\sc vii} lines is not clear, but we think that it is in part 
due to remaining uncertainties in the atomic data, as well as those in the 
flux measurements. A fuller atomic model for the He\,{\sc i}-like ions is 
clearly needed in CHIANTI. Ideally, the value of $R_o$ should be established
 from observations of the {\it quiet} solar corona, where the density is 
expected to be sufficiently low to give this limiting ratio. \cite{gj73} 
used an observed ratio of 3.6 in this manner, although \cite{blum72} quote
 higher ratios of 3.78 and 3.92. Unfortunately, the LETGS spectra of
$\alpha$~Cen A \citep
{ness02}, where a solar-like pressure might be expected,
 do not have sufficient flux in the O\,{\sc vii} lines to measure $R_o$ to
within useful limits.

The {\it EUVE} lines of Fe\,{\sc xiv} are formed around $\log T_{\rm e} = 6.25$,
within the range over which the O\,{\sc vii} lines are formed. \cite{laming96}
found $\log P_{\rm e} = 15.25$ from the lines at 211.33~\AA\ and 219~\AA, but the 
latter is weak and blended. Using the 211.33-\AA\ and 264.78-\AA\ lines, 
\cite{schmitt96} found $\log P_{\rm e} = 15.55$, still lower than that indicated by
 the O\,{\sc vii} and transition region lines. 

A number of other line ratios are sensitive to $n_{\rm e}$, because the
 relative
populations of the levels in the ground term are not given by the Boltzmann
population (e.g. lines in the B\,{\sc i}-like isoelectronic sequence and lines
of iron). The densities derived depend on the overall form of the EMD and 
relative element abundances and are discussed in Section~\ref{S4.5} and 
Section~\ref{S4.6.1}.

In the calculations that follow in Sections~\ref{S4} and ~\ref{S5} we have 
explored the results using pressures of $\log P_{\rm e} = 15.30$, 15.68 and 
16.10. In Section~\ref{S6} we require the theoretical models to produce
a value of $\log P_{\rm e} = 15.97\,\pm\,0.20$ at $\log T_{\rm e} = 5.3$. 
 
\section{Emission measure loci and line identifications}
\label{S4}

The identifications of the strong lines in the LETGS range are well known.
 In identifying other lines, and to check for blends, we used both APED 
(ATOMDB v1.3.1) and the CHIANTI database (v5.2) \citep{landi06} to 
explore which transitions might be present at a given wavelength. Emission 
measure loci (EMLs) were then calculated for possible candidates, including 
any dependence on $P_{\rm e}$.
  
For a spherically symmetric atmosphere, the line flux observed at the Earth 
is given by

\begin{equation}
\label{eq1}
F_{21} =  \frac{R_{*}^{2}}{d^2} \frac{hc}{\lambda} \frac{n_{E}}{n_{\rm H}} 
          \int \frac{n_2}{n_{ion}} 
          \frac{A_{21}}{n_{\rm e}} \frac{n_{ion}}{n_{\rm E}} G(r) f(r) n_{\rm e} n_{\rm H}
          {\rm d}r   % (1) 
\end{equation}
where $G(r)$ is the fraction of photons not intercepted by the star,
$f(r) = r^{2}/R_{*}^2$ and $d$ is the distance to the star. The excited and 
lower levels are $2$ and $1$, respectively (where $1$ is not necessarily the
 ground state); 
$n_{E}/n_{\rm H}$ is the abundance of the element relative to hydrogen,
taken as constant over the region of line formation; $n_{ion}/n_{E}$ is the
relative ion population; $n_{\rm H}$ is the hydrogen number density; $A_{21}$ is the
spontaneous transition probability and the integration is over the radial 
distance, d$r$.

Equation (\ref{eq1}) can be rewritten as
\begin{equation}
\label{eq2}
F_{21} \simeq g(n_{\rm e}, T_{\rm e}) \int n_{\rm e} n_{\rm H} G(r) f(r) {\rm d}r    % (2)
\end{equation}
where $g(n{\rm e}, T_{\rm e})$ includes all other terms in equation (\ref{eq1}).
Provided any dependence on $P_{\rm e}$ is taken into account, the emission
measure locus (EML) gives the value of the {\it apparent} emission measure 
($EM = \int n_{\rm e} n_{\rm H}~G(r)~f(r)$~d$r$)
that would be required to account for all the observed flux, at each value
of $T_{\rm e}$ in turn. The loci therefore provide useful constraints on the 
mean EMD, since if this exceeds the minimum of a locus by more than a small 
factor, too much flux will be predicted when the mean EMD is used to predict
 the line fluxes.  

It is important to note that loci from lines of different isoelectronic
sequences can have different variations of $g(T_{\rm e})$ with $T_{\rm e}$, 
owing to the
systematic differences in $n_{ion}/n_{E}$ as a function of $T_{\rm e}$.
Otherwise, if a line with a broad $g(T_{\rm e})$ function were compared with 
one 
with a narrow $g(T_{\rm e})$ function, an incorrect relative element 
abundance 
would be deduced. These differences are taken into account in finding the 
mean EMD (see Section~\ref{S5}).  

Values of $n_{2} A_{21}/n_{\rm e}$ have been calculated using CHIANTI (v5.2). 
The relative ion populations for iron have been taken from \cite{aray92}.
For O\,{\sc vi}, the calculations given in \cite{simjordan05} have 
been adopted, since these include the density dependence of di-electronic
 recombination. All other values are taken from \cite{arot85}.
The element abundances initially adopted and the corrections to these required
by the observations are discussed in Section~\ref{S5}.    

The lines of elements other than iron are now discussed according to their
isoelectronic sequence. Comparisons have been made between observed and 
predicted line flux ratios (or relative EMLs) using a single temperature
for the line formation and also using the total fluxes predicted using the
final EMD. Unless otherwise stated, both approaches give the same results.
Although only the lines that we regard as the most reliable are used to
determine the mean EMD, we calculate the predicted fluxes in all the lines 
discussed below and later compare these with the observed fluxes.

\subsection{Hydrogen-like lines}
\label{S4.1}

The (unresolved) Lyman $\alpha$ lines of C\,{\sc vi}, N\,{\sc vii}, 
O\,{\sc viii}, Ne\,{\sc x} and Mg\,{\sc xii} are all observed, although 
the Mg\,{\sc xii} lines are weak. The Ne\,{\sc x} lines at 12.13 + 12.14~\AA\ 
are blended with a line of Fe\,{\sc xvii} at 12.12~\AA. Another line of
Fe\,{\sc xvii} is observed at 12.27~\AA. The ratio of these two Fe\,{\sc xvii}
lines does not depend significantly on $T_{\rm e}$ so we have used the line at
12.27~\AA\ to predict the flux in the line at 12.12~\AA. This results in
73~per cent of the observed total flux in the line at 12.13~\AA\ being due to
Ne\,{\sc x}. 

The Lyman $\beta$ lines are observed in O\,{\sc viii} (at 16.01~\AA) and 
Ne\,{\sc x} (at 10.23~\AA), but
the former is blended with a line of Fe\,{\sc xviii} at 16.00~\AA.
Another line of Fe\,{\sc xviii} at 16.07~\AA\ has been used to find the
contribution of the 16.00-\AA\ line to the total flux. The ratio of the
Fe\,{\sc xviii} lines is slightly sensitive to $T_{\rm e}$, and their 
temperature of optimum formation could lie between $\log T_{\rm e} = 6.5$ and 
6.8.
The total predicted fluxes show that 81~per cent of the observed line at 
16.01~\AA\ is due to O\,{\sc viii}. The ratio of the observed flux in
 O\,{\sc viii} Lyman $\beta$ to that in the Lyman $\alpha$ line is then a 
factor of 1.2 larger than that predicted using the mean EMD.   
In Ne\,{\sc x}, the observed Lyman $\beta$ to Lyman $\alpha$ ratio is a 
factor of 1.4 larger than predicted.

\subsection{Helium-like lines}
\label{S4.2}

The resonance (r), intersystem (i) and forbidden line (f) of C\,{\sc v}
lie in a region of low and rapidly varying effective area and a real signal
is observed only in the r line at 40.27~\AA. However, this is slightly
blended with the third order of the Ne\,{\sc ix} r-line and it is hard to
determine a reliable flux \citep{ness01}. The uncertainty in the flux
given in Table~\ref{tab3} includes the results of varying the background and 
line width. The r-line of N\,{\sc vi} is present, although rather weak; the i
 and f-lines are not sufficiently above the local noise level to be useful. 
The lines of O\,{\sc vii} are relatively strong and their use in constraining 
$P_{\rm e}$ has been discussed in Section~\ref{S3}. Note that the contribution
from radiative recombination has been included when predicting the fluxes in 
the singlet lines of all the He\,{\sc i}-like ions.

\begin{figure}
\resizebox{\hsize}{!}{\includegraphics{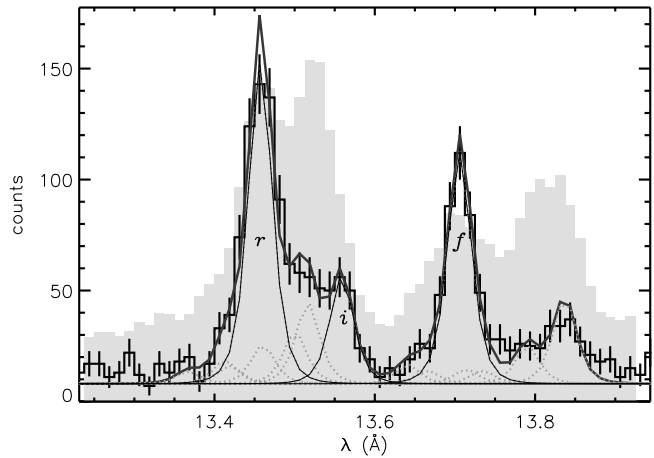}}
\caption{\label{f2}The spectral region around the Ne\,{\sc ix} triplet in 
$\epsilon$~Eri and Capella (light grey shade), rescaled to match the peak
 flux in the r line. The contributions from the lines of Fe\,{\sc xvii}, 
{\sc xviii} and {\sc xix} in $\epsilon$~Eri are shown by dotted lines. Those
 for the r, i and f lines are shown by full lines. The blending in 
$\epsilon$~Eri is clearly far less severe than in Capella. The method of 
fitting is described in the text. The fluxes derived for the Ne\,{\sc ix} 
lines, and their uncertainties, are given in Table 3.}
\end{figure}

The lines of Ne\,{\sc ix} are also relatively strong, but are blended to 
various degrees with lines of Fe\,{\sc xvii}, {\sc xviii} and {\sc xix}.  
A comparison between the LETGS spectra of $\epsilon$~Eri and Capella, shown in 
Fig.~\ref{f2}, indicates that the blends with lines of iron are less
important than in Capella. In $\epsilon$~Eri, the Fe\,{\sc xvii} line at 
13.83~\AA\ is significantly weaker than the Ne\,{\sc ix} f line
and the Fe\,{\sc xix} + {\sc xxi} blend at 13.51~\AA\ is significantly weaker
 than the r line. The de-blending problem is therefore less severe than in
Capella. The procedure used is as follows: the amplitudes and wavelengths of
 the lines observed in Capella using the High Energy Transmission Grating 
Spectrograph (HETGS) \citep[see][]{ness03a} are used to predict the counts 
in the LETGS spectra, taking into account the resolution and effective area of
 the LETGS; the lines of Fe\,{\sc xvii} and Fe\,{\sc xviii} are grouped 
together, as are the lines of Fe\,{\sc xix}, while the lines of Ne\,{\sc ix} 
are treated individually; the amplitudes of the 2 groups and 3 individual 
lines are scaled iteratively to obtain the best fit to the LETGS spectrum of 
$\epsilon$~Eri and hence the fluxes in the Ne\,{\sc ix} lines are determined.
The fits made are shown in Fig.~\ref{f2}. The analyses of all lines of neon 
and other lines of iron shows that the relative weakness of the blended iron
 lines in $\epsilon$~Eri arises from both the lower EMD above 
$\log T_{\rm e} = 6.6$ and a lower Fe/Ne relative abundance.     
Given the de-blending process, in particular for the i-line, it is difficult
 to make a reliable interpretation of the relative fluxes in the Ne\,{\sc ix}
r, i and f lines.  

The r and f lines are present in Mg\,{\sc xi} and Si\,{\sc xiii}, but the i 
lines cannot be distinguished. In extracting fluxes we have summed over the 
r, i and f lines. The predicted fluxes are lower limits, since the 
contribution to the 
populations of excited states from dielectronic recombination is not yet 
included in CHIANTI (v5.2), and contributions from satellite lines are not
yet included for Mg\,{\sc xi}. 

The 1s3p~$^1$P -- 1s$^2$~$^1$S transition in O\,{\sc vii} is present
 as a weak line at 18.63~\AA, but is blended with a weaker line.
 The same transition in Ne\,{\sc ix} might also be present at 11.54~\AA. 
These weak lines are not used in deriving the mean EMD.
 
\subsection{Lithium-like lines}
\label{S4.3}

In the region below 170~\AA\ the transitions observed in the Li\,{\sc i}-like
ions are those between the n = 3 and n = 2 levels (3p -- 2s, 3d -- 2p, 3s --
2p). The 3p -- 2s (unresolved) lines in O\,{\sc vi} at 150.1~\AA\ are present
in the minus direction spectrum, but are barely above the instrumental noise 
level in the plus direction spectrum. The measured flux is very sensitive     
to the background adopted and provides only an upper limit to the EML. 
Although the 3p -- 2s lines are not included in the derivation of the 
mean EMD, this reproduces the observed flux to within a factor of 1.1.
The 2p -- 2s transitions are both observed (around 1032~\AA\ and 1038~\AA)
in spectra obtained with {\it FUSE} in 2000 December. They are useful in
constraining the EMD between $\log T_{\rm e} = 5.5$ and 6.1 (see Section~\ref{S5}).

In Ne\,{\sc viii} the 3p -- 2s and 3d -- 2p transitions are observed at around
88.1~\AA\ and 98.2~\AA, respectively. Although the 3s -- 2p transitions
are just present around 103.1~\AA, they are too weak to yield a reliable flux.
The 4d -- 2p transitions occur around 73.5~\AA\, but on the basis of the
calculated EMLs, the line observed is identified as one of Fe\,{\sc xv}. 
\cite{heroux72} suggested that in the solar spectrum, the lines around 
88.1~\AA\ are blended with lines of Fe\,{\sc xi} and identify a line at 
86.88~\AA\ with the strongest member of the multiplet (at solar densities). 
At the higher value of $n_{\rm e}$ in $\epsilon$ Eri, the line at 86.88~\AA\ should
 still be the strongest member of the multiplet, but is not observed. We have
 rejected the possibility of a second order line of Si\,{\sc xii} at 
88.04~\AA\ because a stronger second order Si\,{\sc xii} line at 88.34~\AA\ 
is not present. Thus significant blending with these Ne\,{\sc viii} lines 
seems unlikely. Note that the relevant lines of Fe\,{\sc xi}
are not yet included in either CHIANTI or APED. \cite{heroux72} also
suggested that both of the lines at 98.11~\AA\ and 98.26~\AA\ are blended
with other lines in the solar spectrum. In $\epsilon$~Eri, to within the
measurement uncertainties, the two lines have the expected wavelength  
separation and almost the theoretical flux ratio, so there is no obvious 
evidence of substantial blending.

The lines in Mg\,{\sc x} are all weak. The 3p -- 2s multiplet at 57.90~\AA\
is not resolved; only the summed flux is used. The 3d -- 2p transitions at 
63.30~\AA\ + 63.31~\AA\ and 63.16~\AA\ are observed only in the minus 
direction spectrum, owing to a chip gap in the plus direction. 
We use only the blend of transitions at around 63.3~\AA\ in deriving the EMD,
 since the weaker line at 63.16~\AA\ is barely above the noise level.  
The 3s -- 2p lines at 65.67~\AA\ and 65.85~\AA\ have an incorrect 
flux ratio in the minus side spectrum and summed spectra, suggesting that the
intrinsically weaker line at 65.67~\AA\ is blended. Also, the line at 
65.85~\AA\ might be blended with another line at around 65.93~\AA\, making 
it difficult to extract a reliable flux; these two Mg\,{\sc x} lines are 
excluded from the derivation of the mean EMD. 

In Si\,{\sc xii} the 3p -- 2s lines lie around 40.93~\AA\, in the region
of the instrumental absorption edge, and are not observable. The 3d -- 2p
transitions at 44.02~\AA\ and 44.17~\AA\ are observed as moderately strong 
lines. The observed ratio of the stronger to weaker components is about 
1.5,  instead of the predicted value of 1.99. We use only the stronger line,
on the grounds that the weaker line might be blended.
The 3s -- 2p lines at 45.52~\AA\ and 45.69~\AA\ are also observed, but are not
used in deriving the mean EMD, as they are weak and blended with other weak 
lines or instrumental noise.  

There are weak lines that are possibly due to S\,{\sc xiv}. A line at 
32.56~\AA\ corresponds to the stronger component of the 3d -- 2p transitions.
Another at 30.46~\AA\ corresponds to the 3p -- 2s transitions, but the
2p$^5$3d~$^1$P -- 2p$^6$~$^1$S transition in Ca\,{\sc xi} occurs at 30.47~\AA\
and could contribute to the measured flux. These lines were not used in
deriving the mean EMD.

In Ne\,{\sc viii} the ratio of the observed flux from the 3p -- 2s 
transitions to that in the 3d -- 2p transitions is about a factor of 1.5 
larger than predicted. This suggests that the atomic models or atomic data 
for the lithium-like ions would bear closer examination.  

\subsection{Beryllium-like lines}
\label{S4.4}

No lines of Mg\,{\sc ix} are observed. Because it is useful to constrain the
EMD at around $10^6$~K, we have used the background level at 72.31~\AA\
and 77.74~\AA\ (the wavelengths of the 2s3d -- 2s2p and 2s3s -- 2s2p
 transitions, respectively), to find the upper limit to their EMLs.

Three singlet lines of Si\,{\sc xi} are observed; the 2s3p -- 2s$^2$  line
at 43.76~\AA, the 2s3d -- 2s2p line at 49.22~\AA\ and the 2s3s -- 2s2p line
at 52.30~\AA\ (but the latter only in the plus side spectrum).  
Only the 52.30-\AA\ line appears to be unblended.

The strongest transition (2s3d -- 2s2p) in S\,{\sc xiii} is observed at 
35.67~\AA; other transitions are present, but are barely above the noise 
level. 

\subsection{Boron-like lines}
\label{S4.5}

The 3d -- 2p lines of Si\,{\sc x} lie at 50.69~\AA\ ($^2$P$_{3/2}$ -- 
$^2$D$_{5/2,3/2}$) and 50.52~\AA\ ($^2$P$_{1/2}$ -- $^2$D$_{3/2}$). 
As in Fe\,{\sc xiv} \citep{jordan65}, the ratio of these lines depends on 
$n_{\rm e}$, since the relative populations of the ground $^2$P levels have a 
Boltzmann distribution only at large values of $n_{\rm e}$. The line at 
50.52~\AA\ is blended with a line of Fe\,{\sc xvi} at 50.55~\AA\ and careful 
deblending is required. Using the procedure described in Section~\ref{S4.6.3},
the ratio of the deblended fluxes in the 50.69 and 50.52-\AA\ lines is $1.52\,\pm\,0.26$. Using CHIANTI (v5.2) and the mean EMD, the predicted integrated 
fluxes give ratios of 1.51 at $\log P_{\rm e} = 15.68$, 1.18 at 
$\log P_{\rm e} = 15.30$ 
and 1.78 at $\log P_{\rm e} = 16.10$. Thus a value of $\log P_{\rm e}$ 
close to 15.68
gives the best fit, but the uncertainty in the flux ratio just includes 
$\log P_{\rm e} = 16.10$. The flux in the unblended Si\,{\sc x} line at 
50.69~\AA\ 
is also weakly dependent on $n_{\rm e}$ (compared with the $n_{\rm e}^{2}$ 
dependence of 
most lines). At $\log P_{\rm e} = 15.68$, the flux predicted by CHIANTI (v5.2) 
is a factor of 1.3 larger than that observed and a pressure of $\log P_{\rm e} \le
15.30$ would be required to fit the observed flux. 
Predicted flux ratios are in general more accurate than absolute fluxes,
since they do not depend on abundances or ion fractions, so there may be small
problems with the atomic data. 
    
The corresponding lines of S\,{\sc xii} are not observed.

\subsection{Lines from iron ions}
\label{S4.6}

We now discuss the lines of iron according to their stage of ionization.

\subsubsection{Fe\,{\sc ix} to Fe\,{\sc xiv}}
\label{S4.6.1}

For these ions 
we rely on the $\Delta n = 0$ transitions observed with the {\it EUVE}, 
although 
the Fe\,{\sc ix} line is also observed with the LETGS. We also discuss the 
Fe\,{\sc xii} forbidden lines that are observed with the STIS. At 
present, neither CHIANTI (v5.2) nor APED include all transitions of the type 
$\Delta n = 1$ in these ions. We have updated earlier calculations of the
 EMLs by using CHIANTI (v5.2) \citep[including those by][who used 
CHIANTI v4]{simjordan03a}.

The lines of Fe\,{\sc ix} to {\sc xiv} used are {\it all} sensitive to 
$n_{\rm e}$,
either through the departure from Boltzmann populations in the levels of
the ground term, or from the population of higher metastable states. Apart from
the blended lines of Fe\,{\sc xiii} at 203.83~\AA\, the derived EMLs all 
increase with increasing $n_{\rm e}$. The stronger 203.83-\AA\ line ends on an 
excited level of the ground term and thus has the opposite behaviour. To 
give a smoothly increasing EMD from all the lines of Fe\,{\sc x} to {\sc xv} 
would require a value of $\log P_{\rm e} \le 15.30$; at higher pressures the 
Fe\,{\sc xiii} locus lies below the mean value. Since there is no other
evidence of such a low pressure, there might be small problems with the atomic
data for Fe\,{\sc xiii} in CHIANTI (v5.2) or in the {\it EUVE} line fluxes. 
With the 
inclusion of more levels in the atomic models of these ions, it might be 
possible to derive a value of $P_{\rm e}$.

The resonance line of Fe\,{\sc ix} at 171.07~\AA\ leads to a relatively 
low EML, irrespective of pressures in the range from $\log P_{\rm e} = $ 
15.3 to 16.1. Even at $\log P_{\rm e} = 16.10$, the mean EMD leads to a 
flux that
is larger than the value observed with the {\it EUVE} by a factor of 1.8.
 The 171-\AA\ line falls near the long wavelength limit of the LETGS and the
short wavelength limit of the {\it EUVE} medium wavelength spectra; 
for both instruments there are significant uncertainties in the flux
calibration at 171~\AA\ and these could be one origin of the above discrepancy.
  As discussed by \cite{laming96}, the ion balance calculations by 
\cite{arot85} lead to a larger EML than that found from the calculations by 
\cite{aray92}. Near the peak emissivity, the difference between these
 calculations is a factor of 1.6, which would remove much of the discrepancy. 
However, the calculations by \cite{aray92} are expected to be more accurate.
The final model can be used to estimate the line centre opacity. This is
close to 1; although scattering of photons out of the line of sight could
occur, detailed radiative transfer calculations are needed to find the
effect on the spatially integrated line fluxes.
 
The atomic data for Fe\,{\sc xii} have been revised since the work by
\cite{jordan01a,jordan01b}, who used CHIANTI (v3.01), and by 
\cite{simjordan03a}, who used CHIANTI (v4.2). We have therefore re-examined 
the difference between the fluxes predicted by CHIANTI (v5.2)
for the forbidden lines at 1242 and 1349~\AA\ and the {\it EUVE} lines in 
the blend around 196~\AA. For the lines at 1242~\AA\ and at around 195~\AA, 
\cite{jordan01a} found a difference of a factor of 3 between their EMLs.
 This is now reduced to a factor of 1.8 (at $\log P_{\rm e} = 15.68$) or 
2.2 and 1.5 (at $\log P_{\rm e} = 15.30$ and 16.10, respectively). 
The small dependence on $P_{\rm e}$ arises from a small increase in the 
forbidden 
line fluxes, and a small decrease in the EUV line fluxes, with increasing 
$P_{\rm e}$. Using the absolute line fluxes and the mean EMD, the agreement
between the observed and predicted fluxes for the EUV lines is very good (to
within a factor of 1.1 over the above range of $P_{\rm e}$) but the flux in 
the line
at 1242~\AA\ is predicted to be smaller than that observed, by the factors
given above. Although differences in the fluxes arising from the different
 dates of the observations cannot be ruled out, neither can small corrections 
to the level populations for the forbidden lines (see below). 
      
The ratio of the fluxes in the Fe\,{\sc xii} forbidden lines at 1242~\AA\
 and 1349~\AA\
is insensitive to $\log P_{\rm e}$ over the range from about 15.0 to 16.0, but
is useful in placing an upper limit on $P_{\rm e}$. In $\epsilon$~Eri the 
observed
ratio is 1.88 ($\pm\,0.2$) (and other main-sequence stars show a similar ratio)
 \citep{jordan01a}. Using a single temperature of $\log T_{\rm e} = 6.15$, 
CHIANTI (v3.01) leads to $\log P_{\rm e} = 15.72$, with an upper limit of 
16.17. CHIANTI (v4.2) leads to $\log P_{\rm e} = 15.53$, with an upper limit 
of 16.07. But CHIANTI (v5.2) leads to a pressure of $\log P_{\rm e} = 
14.21$, which is much lower than the transition region pressure found in 
Section 3. The upper limit is 15.80, which is consistent with the 
transition region  pressure (15.97 $\pm 0.20$), but not with the pressure
of $\log P_{\rm e} = 16.14$ found in the final model at 
$\log T_{\rm e} = 6.15$. 
At present we suggest that the atomic data used in CHIANTI (v3.01) or (v4.2)
give a better fit to the forbidden line flux ratio than do those in 
CHIANTI (v5.2) \citep[see][]{storey05}.
 
The value of the Fe\,{\sc xii} forbidden line flux ratio provides a very 
sensitive test of the
 atomic data for these lines. E.g. \cite{jordan01b} pointed out that the ratio 
of 2.7 predicted by \cite{binello01} could not be correct. It is also of 
interest to compare the population of the 3p$^{3}$~$^2$P$_{1/2}$ level from 
CHIANTI (v5.2) with that predicted empirically by \cite{gj75} on the basis 
of solar observations. At $T_{\rm e} = 1.65 \times 10^6$~K and $n_{\rm e} = 
3 \times 10^8$~cm$^{-3}$, CHIANTI (v5.2) gives a level population (relative 
to that 
of the ion) of $2.9 \times 10^{-4}$, whereas the solar observations led to 
values between 3.3 and 5.1 $\times 10^{-4}$. Thus there is other 
observational support for a larger $^{2}$P$_{1/2}$ population.

Given that the final EMD peaks around the temperature where Fe\,{\sc xv} and
 {\sc xvi} are formed, one might expect lines of Fe\,{\sc xiv} to be present
in the X-ray region. There are four weak lines around 76~\AA\ that are
 also present in the LETGS spectra of Procyon \citep{raassen02}, 
$\alpha$~Cen A and B \citep{raassen03} and Capella \citep{sf04}.
In $\epsilon$~Eri the lines are at 75.91~\AA, 76.04~\AA, 76.13~\AA\ and 
76.53~\AA. We propose that the lines at 76.04~\AA\ and 76.13~\AA\ are due 
to the 3d~$^2$D -- 4f~$^2$F transitions in Fe\,{\sc xiv}, but owing to the
absence of these lines in CHIANTI or APED we cannot check this through
derived EMLs. \cite{raassen03} have also proposed this identification
for lines in $\alpha$~Cen A and B. In Procyon, a line at 75.98~\AA\ may well be
due to Fe\,{\sc x} \citep{raassen02}, but in $\epsilon$~Eri the EMD is 
relatively smaller where such lines are formed. Using our final EMD, none of
the four possible lines of Fe\,{\sc xiv} between 75.69~\AA\ and 76.82~\AA\
are predicted to be observable.
A line of Fe\,{\sc xvi} occurs at 76.50~\AA\ but the next strongest
 member of the multiplet at 76.80~\AA\ is absent (see also Section 4.6.3).

\subsubsection{Fe\,{\sc xv}}
\label{S4.6.2}

The resonance line at 284.2~\AA\ is observed as a strong line with the 
{\it EUVE}. Using our final EMD, the flux
 predicted in this line is a factor of 1.16 larger than that observed. 
Possible sources of uncertainty include the amount of absorption by the ISM and
line opacity effects.

The X-ray spectrum of Fe\,{\sc xv} in a solar flare and Capella has been
discussed by \cite{keenan06} and we make comparisons with predicted flux 
ratios at $\log T_{\rm e} = 6.3$ (near where the EMLs for the Fe\,{\sc xv} 
lines
have their minimum value) and $\log T_{\rm e} = 6.5$ (to allow for the 
increase in
the mean EMD). We also make comparisons with the flux ratios predicted using 
CHIANTI (v5.2) and the mean EMD. We observe only singlet transitions
whose flux ratios do not depend on $n_{\rm e}$. 

The 3s4d~$^1$D -- 3s3p~$^1$P (59.40~\AA) transition is adopted as the standard 
line. The blend at 59.27~\AA\ observed in Capella  by \cite{keenan06} is not
obvious in $\epsilon$~Eri, consistent with their suggestion that it is due
to Fe\,{\sc xvii}. The 59.40-\AA\ line flux is a factor of 1.38 larger 
than that predicted using CHIANTI (v5.2) and the mean EMD.

The 3s4p~$^1$P -- 3s$^2$~$^1$S transition at 52.91~\AA\ is not obviously
blended in $\epsilon$~Eri, unlike the situation in Capella. The observed flux
ratio agrees well with that predicted using CHIANTI (v5.2), and although
the flux ratio from \cite{keenan06} is smaller, it also agrees with that
observed to within the uncertainties.  

The strongest X-ray line is both predicted and observed to be the 
3s4s~$^1$S -- 3s3p~$^1$P transition at 69.68~\AA. In $\epsilon$~Eri
this line is blended with one at around 69.6~\AA\ that does not appear to
be present in Capella (\citep{keenan06}). We note that \cite{kelly87} lists
predicted lines of Fe\,{\sc xiv} in this region. The flux ratio observed for 
the line at 69.68~\AA\ is a factor of about 1.5 lower than that predicted by 
both CHIANTI (v5.2) 
and \cite{keenan06}, which agree well with each other. This ratio is also 
lower than expected in Capella, but agrees with the theoretical value to 
within the uncertainties. 
   
The 3s4f~$^1$F -- 3s3d~$^1$D transition occurs at 73.47~\AA, but is 
potentially blended with a line of Ne\,{\sc viii} at 73.48~\AA.
Interpreting the observed line flux as being due entirely to Ne\,{\sc viii}, 
using 
CHIANTI and the mean EMD gives an observed to predicted flux ratio that is a 
factor of 4.0 too large. Assuming that the predicted Ne\,{\sc viii} flux is 
correct, its contribution can be removed to give an Fe\,{\sc xv} flux of 
$2.04 \times 10^{-14}$~erg~cm$^{-2}$~s$^{-1}$ and a flux ratio of 0.68. The
 temperature sensitivity of the Fe\,{\sc xv} 73.47-\AA\ line is not given 
by \cite{keenan06}
so that of the triplet lines from the same configuration has been used to
find the expected flux ratio at $\log T_{\rm e} = 6.50$. This corrected flux ratio 
agrees well with the flux ratio predicted by \cite{keenan06}, but the flux 
ratio predicted using CHIANTI is larger.
However, the stronger member of the Ne\,{\sc viii} multiplet that should occur
at 73.56~\AA\ is not present with the flux expected from CHIANTI. 
For this reason we have also found the observed flux ratio assuming no
contribution from Ne\,{\sc viii}, and this agrees with that predicted using
CHIANTI, to within the uncertainty.

The predicted and observed flux ratios are summarized in Table~\ref{tab5}. 
We have checked that the uncertainties in the observed ratios also cover the
range of
values derived using a range of background levels. Of the X-ray lines, only 
the observed flux ratio for the line at 69.68~\AA\ 
is discordant with the calculations. On average, the flux ratios predicted 
by \cite{keenan06} are smaller than those predicted using CHIANTI (v5.2). 

At present, the atomic model does not include levels with
$n$ larger than 5; this could be one cause of the remaining differences 
between 
the observed and calculated flux ratios for transitions from the $n = 4$ 
levels. Also, from section 2.3 of \cite{landi06}, the
effects of recombination and ionization on the populations of excited states
have not yet been included for Fe\,{\sc xv}. 
       
% Table 5
\begin{table}
\caption{\label{tab5}Fe\,{\sc xv} line flux ratios, relative to the flux in
the 59.40-\AA\ line$^a$.}
\begin{tabular}{llcc}
\hline
Line     & Keenan et al.     & CHIANTI (v5.2)$^b$ & Observed   \\
(\AA)    &  (2006)           &                    &             \\
\hline
         & $\log T_{\rm e}$        &             &                    \\   
52.91    & 6.3~~  $0.65\,\pm\,0.13$  &   0.81      &  $0.75\,\pm\,0.17$  \\
         & 6.5~~  $0.64\,\pm\,0.13$  &             &                  \\
69.68    & 6.3~~  $3.1\,\pm\,0.6$   &   2.75      &  $1.85\,\pm\,0.31$ \\
         & 6.5~~  $2.7\,\pm\,0.5$   &             &                   \\
73.47    & 6.3~~  $0.97\,\pm\,0.16$  &   1.01      &  $0.68\,\pm\,0.20$$^c$\\
         & 6.5~~  $0.75\,\pm\,0.12$  &             &  $0.91\,\pm\,0.22$$^d$ \\
284.2    &                         &   56.0      &  $34.8\,\pm\,5.6$  \\
\hline
\end{tabular}
\\
$^a$ This is $2.99\,\pm\,0.44 \times 10^{-14}$~erg~cm$^{-2}$~s$^{-1}$, when 
corrected for absorption in the ISM. \\
$^b$ Predicted using the mean EMD (see Table 10), using $\log P_{\rm e} = 
16.10$. \\
$^c$ With the predicted contribution from Ne\,{\sc viii} removed. \\
$^d$ Assuming no contribution from Ne\,{\sc viii}. 
\end{table}

\subsubsection{Fe\,{\sc xvi}}
\label{S4.6.3}

In addition to the 3p -- 3s transitions observed with the {\it EUVE}, lines
from the 4p -- 3s, 4d -- 3p, 4s -- 3p and 4f -- 3d transitions are observed
in the LETGS spectrum. Thus it is possible to test the atomic
data used in CHIANTI (v5.2); these were not updated from those used in
CHIANTI (v4.2).   

To make comparisons between the observations and the theoretical values
from \cite{cornille97} and CHIANTI (v5.2), we use the 4s -- 3p line at
 63.72~\AA\ as the standard line. Using CHIANTI (v5.2) and the integrated
fluxes, the predicted and observed ratios of this and the lines at 335~\AA\
and 361~\AA\ agree to within 7~per cent.
   
The 4p -- 3s transitions at 50.35 and 50.55~\AA\ are blended with the 
third-order of the Fe\,{\sc xvii} line at 16.78~\AA\ and with the
density-sensitive line of Si\,{\sc x} at 50.52~\AA. 
The plus-side spectrum was used to carry out the deblending, because the 
effective area varies rapidly in the minus-side spectrum. The fluxes in the 
third-order lines of Fe\,{\sc xvii} at 17.05 and 17.10~\AA\ have been 
measured and the ratio of the fluxes in these lines to that at
 16.78~\AA\ have been found from the first-order 
spectrum. Using integrated fluxes it is found that the Fe\,{\sc xvi} line 
at 50.35~\AA\ contributes 71~per cent of the observed flux. At 
$\log T_{\rm e} = 
6.5$, the theoretical ratio of the Fe\,{\sc xvi} lines at 50.35 and 
50.55~\AA\ is 1.93, from \cite{cornille97} or 1.82, from CHIANTI (v5.2),
using integrated fluxes. Hence, using CHIANTI (v5.2), Fe\,{\sc xvi} 
contributes 51~per cent of the line at 50.54~\AA. 
The discrepancy between the flux ratio for the 50.35-\AA\ 
line is then a factor of 1.7, using \cite{cornille97}, or 2.5, using
CHIANTI (v5.2). These factors are significantly larger than the average
found from other lines and these lines are not used in finding the mean EMD. 
 
The 4d -- 3p lines at 54.72 + 54.76 and 54.14~\AA\ are observed only
in the plus-side spectra, but are relatively strong, clean lines. 
Using the calculations by \cite{cornille97}, the flux ratio for the 
blended lines at $\simeq 54.74$~\AA\ is a factor of 1.3 smaller than that 
observed. Using CHIANTI (v5.2), at the same $T_{\rm e}$ or using integrated 
fluxes,
the predicted flux ratio is a factor of 1.9 smaller than that observed.   

% Table 6
\begin{table}
\caption{\label{tab6}Fe\,{\sc xvi} line flux ratios, relative to the flux in 
the 63.72-\AA\ line$^a$.
}
\begin{tabular}{lccc} 
\hline
Line    & Cornille et al.&    CHIANTI (v5.2)$^b$  &    Observed \\
 (\AA)  &      (1997)           &                    &              \\
\hline
        & $\log T_{\rm e} = 6.5$       &                   &               \\
50.35   &    0.33                &     0.23      &  $0.57\,\pm\,0.08$ \\
50.55   &    0.17                &     0.13      &  $0.31\,\pm\,0.04$ \\
54.14   &    0.29                &     0.21      &  $0.46\,\pm\,0.07$  \\
54.74   &    0.57                &     0.41      &  $0.77\,\pm\,0.10$  \\
62.87   &    0.49                &     0.49      &  $0.44\,\pm\,0.08$  \\
66.26   &    0.41                &     0.34      &  $0.61\,\pm\,0.10$ $^c$ \\
66.37   &    0.58                &     0.51      &  $0.95\,\pm\,0.13$ $^c$\\
335.4  &                        &     14        &    $14\,\pm\,2$      \\
360.8  &                        &      6.9      &  $7.3\,\pm\,1.0$    \\
\hline
\end{tabular}
\\
$^a$ This is $6.25\,\pm\,0.57 \times 10^{-14}$~erg~cm$^{-2}$~s$^{-1}$, when
corrected for absorption by the ISM.\\
$^b$ Predicted using the mean EMD (see Table 10), using $\log P_{\rm e} = 
16.10$. \\
$^c$ With no correction for the upper limit to the contribution from
the 3$^{rd}$ order O\,{\sc vii} 22.10-\AA\ line.
\end{table}

The 4f -- 3d transitions occur at 66.37 and 66.26~\AA. The observed ratios of
the fluxes in these lines to that of the 63.72-\AA\ line are larger than
those predicted by either CHIANTI (v5.2) (using the mean EMD) or
\cite{cornille97}, but the flux ratios using the latter are closer to those
 observed. 
The third order line of O\,{\sc vii} at 22.10~\AA\ occurs between these lines,
but, on the basis of the upper limit to the flux in the O\,{\sc vii} line at 
21.60~\AA, the contribution from the 22.10-\AA\ should be very small and its
inclusion would not remove the above discrepancy. 
 
The 4p -- 3d transitions around 76.5~\AA\ have smaller branching
ratios than the 4p -- 3s transitions. If the weak line at 
76.53~\AA\ is interpreted as being due to Fe\,{\sc xvi}, then the observed 
flux is about a factor of 5 larger than that predicted by CHIANTI (v5.2).
 \citep{cornille97} do 
not give theoretical relative intensities for the 4p -- 3d transitions.
 This large discrepancy rules out the identification of the line
 at 76.5~\AA\ as Fe\,{\sc xvi}, particularly since the next strongest member
 of the multiplet expected at 76.80~\AA\ is not observed. The calculations 
of the transition probabilities by \cite{cornille97} and by 
\cite{eissner99} are in good agreement with each other for these
 transitions.
(We note that the $\omega f$ value for the 4p~$^2$P$_{3/2}$ -- 
3d~$^2$D$_{3/2}$
 transition given by \cite{cornille97} is a factor of 100 larger than 
expected from their $A$-value, and that this typographical error has been 
reproduced in the comparisons made by \citealt{eissner99}.)

The above comparisons are summarized in Table~\ref{tab6}. Again, we have 
checked that the uncertainties in the observed fluxes cover the range of
values found using different background levels. 

Because we have used the line
at 63.72~\AA\ as the standard, the discrepancies would of course all be 
smaller if the predicted flux for this line (and for the 3p --- 3s lines 
observed with the {\it EUVE}) were smaller than indicated by the given 
statistical uncertainties, but this is not supported by
the overall behaviour of the other iron lines. The atomic model and data for 
Fe\,{\sc xvi} have received close attention and the differences between the
 flux ratios predicted using \cite{cornille97} and CHIANTI (v5.2) are 
smaller than those between either of these sources and the observed flux 
ratios. Overall, the flux ratios predicted using \cite{cornille97} at 
$\log T_{\rm e} = 6.50$ are slightly closer to the observed ratios than 
are those 
using CHIANTI (v5.2) and the mean EMD. Relative to the 4s -- 3p and 4f -- 3d
transitions, the other line flux ratios do increase with $T_{\rm e}$, but 
the mean 
EMD shows that little material exists at the very high temperatures required to
bring the predicted ratios closer to those observed.  Also, the lower 
fluxes measured for the higher ions of iron with the LETGS, compared with 
those measured from {\it EUVE} or the RGS make it unlikely that flaring was 
present in the LETGS spectra.

As for Fe\,{\sc xv}, the effects of including further $n$-states and
 of recombination and ionization to and from excited states still need to be 
investigated. 

\subsubsection{Fe\,{\sc xvii}}
\label{S4.6.4}

The 2p$^5$3d -- 2p$^6$ transitions are observed at 15.01~\AA\ and 15.26~\AA\
 and the 2p$^5$3s -- 2p$^6$ transitions are observed at 16.78~\AA\ and 
17.05 + 17.10~\AA. The predicted flux in the 15.01-\AA\ line is only slightly 
too large, in spite of the fact that the lower collisional excitation rate
suggested by laboratory measurements by \cite{brown06} has not been adopted
 in CHIANTI (v5.2). The line at 15.26~\AA\ is observed to be a factor of 
1.45 stronger than predicted. Without knowing $n_{\rm e}$ at high values of
$T_{\rm e}$, optical depth effects cannot be estimated. However,
\cite{brschm06} have 
suggested that an inner shell transition of Fe\,{\sc xvi} occurs at 
15.26~\AA, but its contribution is not included in the predicted flux. 

\begin{table*}
\begin{flushleft}
\caption{\label{tab7}Photospheric abundances relative to solar values, with
their $\pm$ uncertainties, where available.}
\begin{tabular}{p{.5cm}ccccccp{.9cm}r}
\hline
[C/H] & [O/H] & [Mg/H] & [Si/H]     & [S/H] & [Ca/H] & [Fe/H] & [Ni/H] &
Reference \\
\hline
      &       &        & -0.10 (0.05)&    & -0.10 (0.07)& -0.07 &\mbox{-0.16 (0.04)}&
\cite{bodaghee03} \\
      &       &        &            &     & -0.10 (0.05)& -0.09 (0.05)&   &
\cite{drakesmith93} \\
\mbox{-0.06} &-0.16 (0.02)&-0.14 (0.05)&-0.16 (0.02)&-0.01 (0.01)&-0.11 (0.03)
&-0.12(0.01)& \mbox{-0.20 (0.03)} & \cite{zhao02} \\
\mbox{-0.24} &-0.04  & -0.03  & -0.01      &     & -0.01       & -0.06       & -0.06&
\cite{ap04} \\
\hline
\end{tabular}
\end{flushleft}
\end{table*}

\subsubsection{Fe\,{\sc xviii}} 
\label{S4.6.5}

The resonance lines of Fe\,{\sc xviii} at 93.92 and 103.94~\AA\ are observed
with the {\it EUVE} and the LETGS, but the latter line is weak in both spectra.
Since the ratio of its flux to that of the line at 93.92~\AA\ does not agree 
with the theoretical value, the line at 103.94~\AA\ is not used in 
determining the EMD.   

Fe\,{\sc xviii} has many transitions of the type $\Delta n = 1$ so there
are few strong lines. The optimum temperature of line formation in a uniform
plasma is $\log T_{\rm e} \simeq 6.8$, whereas the mean EMD peaks at around 
$\log T_{\rm e} = 6.6$. Thus the lines of Fe\,{\sc xviii} are relatively weaker
than in stars that have hotter coronae. 

As discussed in Section~\ref{S4.1}, the O\,{\sc viii} Lyman~$\beta$ line at 
16.01~\AA\ is blended with a transition of Fe\,{\sc xviii} at 16.00~\AA.  
The observed relative fluxes of the strongest lines at 14.21 + 14.26~\AA, 
16.08~\AA\ and 93.92~\AA\ agree with those predicted by CHIANTI (v5.2) and
the mean EMD to within a factor of 1.1. 

\subsubsection{Fe\,{\sc xix} and {\sc xx}}
\label{S4.6.6}

The observed lines of these ions lie between 100 and 130~\AA\, where 
\cite{sf04} point out that there are suspected problems with
the flux calibration. The noise levels are also large. 
 
The line of Fe\,{\sc xix} at 108.36~\AA\ is weak, and that at 101.55~\AA\ 
is barely above the noise level. The EML from the line at 108.4~\AA\ shows 
that the mean EMD decreases rapidly at temperatures above about 
$\log T_{\rm e} = 6.7$. The observed ratio of these two lines (which has only
 a very 
small dependence on $n_{\rm e}$ over the range of interest) is larger than 
predicted,
probably because of difficulty in extracting a reliable flux for the weaker 
line.
 
The decrease in the EMD at higher temperatures is confirmed by the line of 
Fe\,{\sc xx} at 132.8~\AA. Lines from higher ions are not definitely observed.
 Lines that are formed above the mean coronal temperature (allowing for the 
extensions of their EMLs to lower temperatures) are likely to be formed in 
stellar active regions, by analogy with the behaviour of the solar EMD. 
Since we do not have values of $n_{\rm e}$ at these high temperatures, we 
cannot model the active region component. 

\section{Relative abundances and the mean emd}
\label{S5}

\subsection{Relative photospheric abundances}
\label{S5.1}

The photospheric element abundances derived for $\epsilon$~Eri by
\cite{abia88,drakesmith93, zhao02,bodaghee03} (and references concerning 
Fe therein), were discussed by \cite{simjordan05}, in the context of relative 
abundances in the lower transition region. All find photospheric abundances of
 iron that are lower than the solar value. 
 \cite{abia88} used early photospheric models, and the solar abundances they
derive differ significantly from recent values; their results are not
considered further here. The abundances derived by the other authors, relative
to solar values, are given in Table~\ref{tab7}. These include the more recent
values from \cite{ap04} that were adopted by \cite{wood06}. 
Both \cite{bodaghee03} and \cite{zhao02} used LTE model atmospheres by 
\cite{kurucz93}. 
\cite{bodaghee03} used the solar abundances by \cite{andersgrevesse89} in
their solar models, but  \cite{zhao02} determined the 
differential abundances using their observations of the Moon, so their 
results have quite small uncertainties. \cite{ap04} also determined
differential abundances using the same lines in the solar spectrum.
 The set of values by \cite{zhao02} 
is the most complete and we have adopted these as the initial values in 
deriving the EMLs, but also make comparisons with those by \cite{ap04}.
 \cite{sf04} also made comparisons with the relative abundances by 
\cite{zhao02}. To
 convert the differential photospheric abundances in $\epsilon$~Eri to 
absolute abundances we have used the solar photospheric abundances 
recommended by \cite{asplund05}. While the solar photospheric abundances 
adopted affect the stellar photospheric abundances, they do not influence the
relative stellar coronal abundances discussed in Section 5.3. The solar 
abundances recommended by \cite{asplund05} are based on exploratory 
3-dimensional modelling and are not directly comparable with the results of the
1-dimensional stellar photospheric models. When comparing the stellar 
photospheric and coronal abundances in Section 5.3, we have also investigated 
results using solar abundances from \cite{grevesse98} that are also based on
1-dimensional modelling. Photospheric abundances are not available for
nitrogen or neon. For nitrogen the mean of the carbon and oxygen differential 
abundances was used, whereas for neon, the solar abundance from 
\cite{asplund05} was initially adopted. 

\subsection{Derivation of the mean EMD}
\label{S5.2}

\begin{figure}
\resizebox{\hsize}{!}{\includegraphics{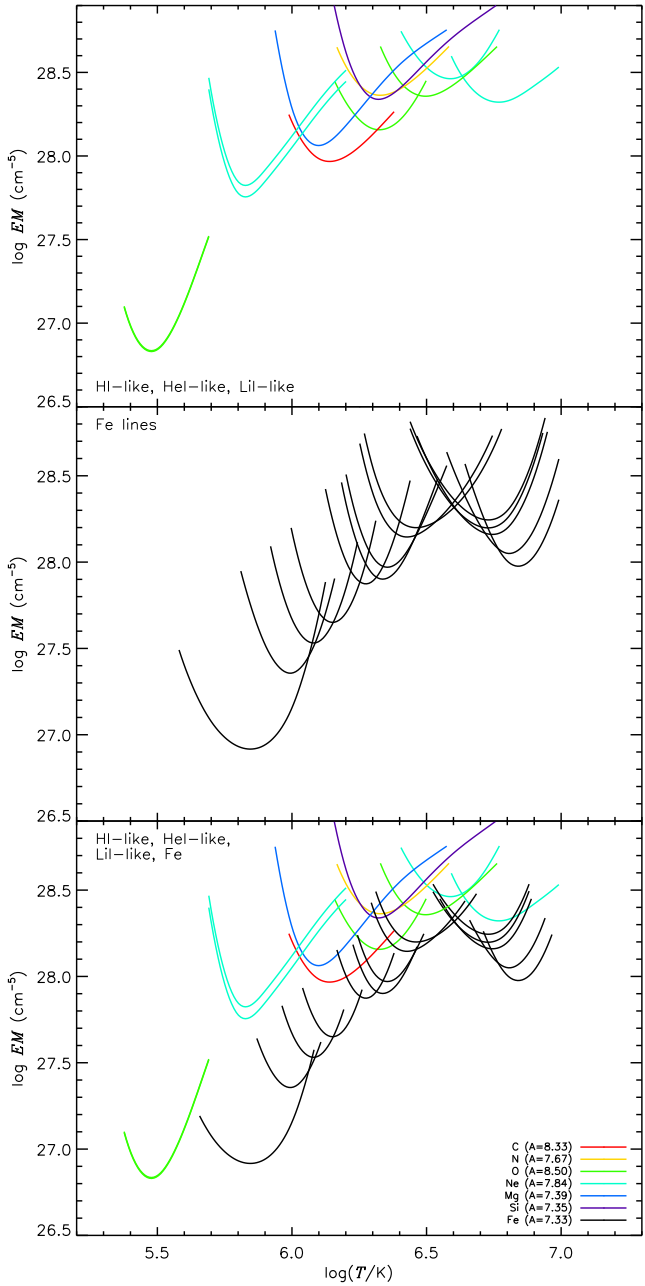}}
\caption{\label{f3}Emission Measure Loci for lines from the
H\,{\sc i}-like, He\,{\sc i}-like and Li\,{\sc i}-like
isoelectronic sequences (top panel), lines of iron (middle panel), and
all these lines (bottom panel). The photospheric abundances initially 
adopted for the transition region and corona of $\epsilon$~Eri are given in
 the bottom right legend. A value of $\log P_{\rm e} = 16.10$ was adopted. The
EMLs are apparent values - see Section~\ref{S6}.}
\end{figure}

\begin{figure}
\resizebox{\hsize}{!}{\includegraphics{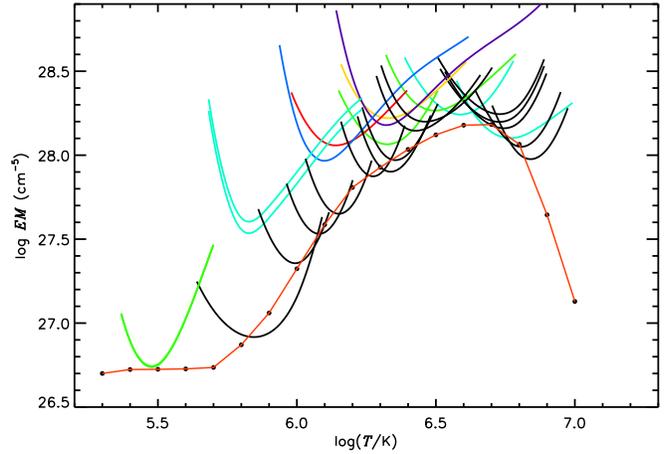}}
\caption{\label{f4}Emission Measure Loci for the most reliable lines,
and the best-fit mean apparent emission measure ($EM(0.3)_{\rm app}$ - see
Section~\ref{S6}) (solid line with grid points). Here the final coronal
abundances are used - see Table~\ref{tab8}. A value of $\log P_{\rm e}
= 16.10$ was adopted.}
\end{figure}

Fig.~\ref{f3} shows the EMLs for lines from the H\,{\sc i}-like, 
He\,{\sc i}-like and Li\,{\sc i}-like isoelectronic sequences (top panel),
the EMLs from the lines of iron (middle panel) and all these lines (bottom 
panel). A value of $\log P_{\rm e} = 16.10$ was adopted. These EMLs are 
apparent values, as defined in Section~\ref{S6}. Apart from the 
lines of Fe\,{\sc ix}, {\sc xvi} and {\sc xvii},
the iron lines are formed over much smaller ranges of $T_{\rm e}$ than those
from the above isoelectronic sequences. Because of these differences the 
impression that the relative abundance of iron adopted is 
{\it significantly} too large is not correct. However, from the top panel it
appears that the adopted relative abundance of nitrogen and silicon is 
quite accurate, that the relative abundance of carbon and magnesium is too 
large and that the relative abundance of nitrogen and oxygen is too small.
From the bottom panel it appears that the relative abundance of neon and iron
is too small.  

To adjust the relative abundances, the mean EMD must be found and used to
re-calculate the line fluxes; systematic differences in the calculated and 
observed fluxes for lines from the various elements can then be investigated.
When attributing any differences to the effects of abundances, rather than to
the shape of the EMD, it is assumed that the corrections to the abundances 
do not depend on $T_{\rm e}$.   

The mean EMD to be derived is defined in terms of the value of the emission
measure for a logarithmic temperature range of 0.30~dex, hereafter, $EM(0.3)$.
The value of $P_{\rm e}$ is assumed to be constant with $T_{\rm e}$, and the 
results using
 the three values $\log P_{\rm e} = $ 15.68, 15.30 and 16.10 have been 
investigated.

In order to find an initial EMD the differences between the functions
$g(n_{\rm e}, T_{\rm e})$ for the lines used must be taken into account; 
some lines
are formed over a small range of $T_{\rm e}$, while others have contributions
 from
a wide range of $T_{\rm e}$. The procedure developed by \cite{jordanwilson71}
and applied, with some modification, by \cite{griffiths98} is adopted. 
First, the fraction of the line formed over a temperature range of 
$\log T_{\rm e}
= \log T_{m}\,\pm\,0.15$ is calculated, where $T_{\rm m}$ is the temperature at which 
the line emissivity has its maximum value, {\it without} allowing for any
variation of the EMD with $T_{\rm e}$. The total contribution is found by
integrating $g(n_{\rm e} T_{\rm e})$ as a function of $T_{\rm e}$, cutting
 off the 
integration when the value is 0.01 times the maximum value of $g(n_{\rm e} 
T_{\rm e})$.
Thus each locus of possible values of $EM(T_{\rm e})$ is replaced by one value,
 referring to the range $\Delta \log T_{\rm e} = 0.3$ about $\log T_{\rm m}$. 

From the values of $EM(0.3)$ derived from 
each line it is then already clear that, relative to the lines of iron, 
the points for Ne\,{\sc viii} lie above the mean, i.e. the abundance
of neon relative to iron is too small. At this point the de-blended fluxes of 
the lines of Ne\,{\sc x} and Fe\,{\sc xvii} discussed in Section~\ref{S4.1}
were used to derive a new starting abundance of 8.01 for neon. (See details in
Section~\ref{S5.3}.) The individual values from each line have then been 
used to define the initial mean distribution of $EM(0.3)$ with $T_{\rm e}$.

This initial mean distribution, defined at intervals of 0.1 in 
$\log T_{\rm e}$, 
is then iterated, taking into account the $g(n_{\rm e}, T_{\rm e})$ 
functions over 
the full range of $T_{\rm e}$ used, and the effects of the variation of the
 EMD with $T_{\rm e}$ are now taken into account. The ratio of the 
observed to predicted fluxes is found and the iteration continued until
these ratios do not change by more than 1~per cent. 

The total flux in line $j$ at step $k$, $F_{j,k}$, is given by
\begin{equation}
\label{eq3}
F_{j,k} = \Sigma_{i} {EM}(0.3)_{i,j,k} g_{j}(T_{i}) \frac{0.1}{0.3} % 3
\end{equation}
where $i$ covers the temperature range over which the emissivity decreases to 
0.01 of its maximum value and the notation for the $g$ function has
been abbreviated to $g_{j}(T_{i})$. 

At step $k$, the value of $\chi^{2}$, that indicates how good the fit is, is 
defined as
\begin{equation}
\label{eq4}
\chi_{k}^{2} = \Sigma_{j} \frac{(F_{j,k} - F_{j,{\rm obs}})^2}{F_{j,{\rm err}}^2}   % 4   
\end{equation}
where $F_{j,k}$ is the flux predicted for line $j$, 
$F_{j,{\rm obs}}$ is the observed flux in line $j$ and $F_{j,{\rm err}}$ is the error in 
the measured flux of line $j$. 

The correction factor $CF_{j,k}$ required to bring the predicted flux
into agreement with that observed in line $j$ is defined as
\begin{equation}
\label{eq5}
CF_{j,k} = \frac{F_{j,k}}{F_{j,{\rm obs}}}~.     % 5
\end{equation} 

This overall correction must be shared between the various values of
$T_i$ and so a weighted function,
\begin{equation}
\label{eq6}
CF_{i,j,k} = \frac{F{j,k}}{F_{j,{\rm obs}}} \frac{{EM}(0.3)_{i,k} g_{j}(T_{i})}
             {\Sigma_{i} {EM}(0.3)_{i,k} g_{j}(T_{i})}     % 6
\end{equation}
is adopted. This allows
for the differences between the $g_{j}(T_i)$ functions of the different
lines and the variation of $EM(0.3)$ with $T_i$. The 
sum of the values of $\log CF_{i,j,k}$ is then found, including all lines 
used at a given $i$. Between each step, the values of $EM(0.3)_{i,k}$ are 
then corrected using
\begin{equation}
\label{eq7}
\log {EM}(0.3)_{i,k+1} = \log {EM}(0.3)_{i,k} + 
         \frac{\Sigma_{j} \log CF_{i,j,k}}{n_i}~, % 7
\end{equation}
where $n_i$ is the number of lines included at $T_i$.

When the iteration process was completed, the systematic behaviour of the 
ratios of the observed and predicted fluxes for lines from the various 
elements were examined and small adjustments to the initial relative 
abundances adopted were made. We have found values relative to that of iron 
\citep[as did][]{sf04}. The absolute values of the mean EMD will 
depend on the abundance of iron adopted. \cite{sf04} and \cite{wood06} 
derived absolute 
abundances of iron by measuring the line to continuum flux ratio, but we 
considered 
 the continuum to be too weak for this to be useful; the error on the 
absolute abundance of iron found by \cite{sf04} is indeed quite large 
($7.20\,\pm\,0.2$). \cite{wood06} find a value of 7.35, very close to
the value of 7.33 that we adopt. The mean EMD derived using the final 
relative abundances and $\log P_{\rm e} = 16.10$ is shown in 
Fig.~\ref{f4} by the solid line with the grid
 points. Fig.~\ref{f4} also includes the loci for lines of isoelectronic 
sequences omitted from Fig.~\ref{f3} for the sake of clarity. 

The mean EMD has also been derived using $\log P_{\rm e} = 15.30$ and 15.68, 
adopting the above relative abundances. These agree with that derived using
$\log P_{\rm e} = 16.10$ to within mean values of -0.03 and -0.02~dex, 
respectively. The largest
differences occur between $\log T_{\rm e} = 6.0$ and 6.2, where they
are -0.07~dex 
(for $\log P_{\rm e} = 15.30$) and -0.04~dex (for $\log P_{\rm e} = 15.68$).
Thus using a different value of $P_{\rm e}$ has an effect on the resulting
mean EMD that is smaller than the uncertainties arising from the line flux
measurements and the atomic data. 

Alternatively, the relative abundances can be adjusted using the EMDs found
with $\log P_{\rm e} = 15.30$ and 15.68. For a fixed abundance of
iron, these are on average 0.04~dex and 0.02~dex
larger, respectively, than those derived using $\log P_{\rm e} = 16.10$. 
Thus the abundances derived (see Section 5.3) do not depend significantly
on the value of $P_{\rm e}$ used in deriving the mean EMD.  

\subsection{Relative coronal abundances}
\label{S5.3}

In Table~\ref{tab8} we give the stellar photospheric abundances according to 
\cite{zhao02}, scaled from the solar photospheric abundances of 
\cite{asplund05}. The coronal abundances derived from the tables in
\cite{sf04} and \cite{wood06} are also given. 
 The values that we derive are given in the final column, on a scale
where the coronal abundance of iron is set 
equal to the stellar photospheric value. We have investigated the likely 
uncertainties in the coronal abundances in several ways. Because we determine
these relative to a fixed value of the iron abundance, using the full set of
maximum observed fluxes, or the full set of minimum observed fluxes, has very
little affect on the abundances derived. Instead, we give the uncertainties 
that result when the maximum fluxes for the lines of iron are used, together 
with the minimum fluxes of all other lines, and vice-versa.

\begin{table}
\caption{\label{tab8}Stellar photospheric abundances from Zhao et al. (2002),
using Asplund et al. (2005) for the absolute scale; stellar coronal
abundances derived by Sanz-Forcada et al. (2004), Wood \& Linsky
(2006) and from the present work (the latter being scaled to the
stellar photospheric iron abundance).}
\begin{tabular}{p{.4cm}cccr}
\hline
\mbox{Ele-} & Zhao & Sanz-Forcada & \mbox{Wood \&} & This work$^b$\\
ment    & et al. & et al. (2004) & Linsky & \\
& (2002)$^a$ &     & (2006)$^b$   &      \\
\hline
C       &   8.33     &  $8.28\,\pm\,0.18$ & 8.15$^{+0.04}_{-0.06}$&
8.24$^{+0.09}_{-0.07}$     \\
N       & (7.67)$^c$ &  $7.74\,\pm\,0.14$ & 7.71$^{+0.04}_{-0.09}$&
7.82$^{+0.08}_{-0.07}$     \\
O       &   8.50     &  $8.53\,\pm\,0.04$ & 8.46$^{+0.02}_{-0.03}$&
8.59$^{+0.04}_{-0.04}$     \\
Ne      & (7.84)$^c$ &  $8.09\,\pm\,0.08$ & 8.02$^{+0.04}_{-0.02}$&
8.06$^{+0.09}_{-0.06}$     \\
Mg      &   7.39     &  $7.43\,\pm\,0.10$ & \mbox{$7.46\,\pm\,0.07$} &
7.49$^{+0.13}_{-0.13}$     \\
Si      &   7.35     &  $7.34\,\pm\,0.07$ & \mbox{$7.33\,\pm\,0.06$}& 7.51$^{+0.07}_{-0.07}$  \\
S$^d$   &   7.13     &  $7.21\,\pm\,0.15$ & 7.15$^{+0.14}_{-0.26}$&
7.29$^{+0.11}_{-0.15}$     \\
Ca$^d$  &   6.20     &  $6.59\,\pm\,0.20$ & 6.70$^{+0.25}_{-1.01}$&
(6.82$^{+0.14}_{-0.16}$)$^e$\\
Fe      &   7.33     &  $7.20\,\pm\,0.20$ & 7.35                   &[7.33]   \\
Ni$^d$  &   6.13     &  $6.24\,\pm\,0.11$ & 5.98$^{+0.19}_{-0.62}$&
6.14$^{+0.08}_{-0.27}$   \\
\hline
\end{tabular}
$^a$ The likely errors in the relative abundances are as given in
 Table~\ref{tab7}. \\
$^b$ The errors are for the abundances relative to that of iron. \\
$^c$ Not given by Zhao et al. (2002); values initially adopted here - see
text. \\
$^d$ Not used in determining the mean EMD.\\
$^e$ The line involved is too weak to derive a reliable value. \\
\end{table}

From Table~\ref{tab8} it can be seen that on the absolute scale adopted,
all our abundances, except that for silicon, agree with those
 derived by \cite{sf04} to within our joint uncertainties.
 If their abundances are scaled to the photospheric 
abundance of iron (7.33), then all our abundances of elements used in 
finding the mean EMD agree with theirs, to within the 
joint uncertainties. Similarly, apart from those for oxygen and silicon, 
the abundances that we derive agree with those found by \cite{wood06}. 

Although we do not include the possible lines of 
Ni\,{\sc xii} in deriving the mean EMD, we give the mean abundance that 
results, since nickel is a low FIP element. The lines of sulphur are
all weak, and possibly blended, and individually give discordant
abundances; the mean value is given. The possible 
line of Ca\,{\sc xii} is predicted to be a factor of 4.2 weaker than that 
observed, and given the behaviour of the other low FIP elements there
may be problems with the identification, the calibration or the atomic data.  
The possible blend between the Ca\,{\sc xi} and S\,{\sc xiv} lines at 
30.45~\AA\
does not appear to support such a large abundance of calcium. Using the 
observed to predicted fluxes for the other two lines of S\,{\sc xiv} suggests 
an abundance of calcium of 6.46. We do not include calcium in the discussions 
below.

Given that the starting abundances used by ourselves,
\cite{sf04} and \cite{wood06} are different, and that we have used more 
up-to-date atomic data 
and independent flux measurements, these comparisons show that a line-based 
approach to the analysis of X-ray data gives reproducible results. 

However, the mean EMD that we derive differs from
 that found by \cite{sf04}. Our EMD increases smoothly up to a peak
value at around $\log T_{\rm e} = 6.6$, whereas that found by \cite{sf04}
has two peaks, at $\log T_{\rm e} = 6.3$ and $\log T_{\rm e} = 6.75$
 -- 6.8.

\begin{table*}
\renewcommand{\arraystretch}{1.1}
\caption{\label{fluxpreds}
The measured fluxes and the ratios of measured to predicted fluxes.}
\begin{flushleft}
{\small
\begin{tabular}{lllr|lllr}
\hline
Ion$^a$ & $\lambda$ & $F_{\rm meas}^b$ & $\frac{F_{\rm meas}}{F_{\rm pred}}^c$ & Ion$^a$ & $\lambda$ & $F_{\rm meas}^b$ & $\frac{F_{\rm meas}}{F_{\rm pred}}^c$\\
\hline
C\,{\scriptsize V} & 40.27 & 3.35 & $1.14\,\pm\,0.26\,$ & Fe\,{\scriptsize IX} & 171.07 & 14.5 & $0.39\,\pm\,0.04\,$\\
{\bf C\,{\boldmath \scriptsize{VI}}} & 33.74 & 17.1 & $1.03\,\pm\,0.06\,$ & {\bf Fe\,{\boldmath \scriptsize{IX}}} & 171.07$^{d}$ &21.1 & $0.56\,\pm\,0.10\,$\\
N\,{\scriptsize VI} & 28.79 & 3.56 & $1.05\,\pm\,0.16\,$ & {\bf Fe\,{\boldmath \scriptsize{X}}} & 174.53$^{d}$ &20.5 & $1.07\,\pm\,0.19\,$\\
{\bf N\,{\boldmath \scriptsize{VII}}} & 24.78 & 10.8 & $1.03\,\pm\,0.08\,$ & {\bf Fe\,{\boldmath \scriptsize{XI}}} & 180.41$^{d}$ &18.8 & $1.08\,\pm\,0.21\,$\\
O\,{\scriptsize VI} & 150.12 & $<$2.3 & $<$1.0 & {\bf Fe\,{\boldmath \scriptsize{XII}}} & 193.67$^{d}$ &56.4 & $1.05\,\pm\,0.42\,$\\
{\bf O\,{\boldmath \scriptsize{VI}}} & 1031.9$^{e}$ &45.9 & $1.04\,\pm\,0.10\,$ & Fe\,{\scriptsize XII} & 1242.0$^{f}$ &0.98 & $1.54\,\pm\,0.11\,$\\
{\bf O\,{\boldmath \scriptsize{VI}}} & 1037.6$^{e}$ &22.6 & $1.03\,\pm\,0.10\,$ & Fe\,{\scriptsize XII} & 1349.4$^{f}$ &0.52 & $1.81\,\pm\,0.52\,$\\
{\bf O\,{\boldmath \scriptsize{VII}}} & 21.60 & 41.5 & $0.98\,\pm\,0.04\,$ & Fe\,{\scriptsize XIII} & 203.83$^{d}$ &19.9 & $0.51\,\pm\,0.19\,$\\
O\,{\scriptsize VII} & 21.81 & 9.59 & $1.28\,\pm\,0.06\,$ & {\bf Fe\,{\boldmath \scriptsize{XIV}}} & 211.32$^{d}$ &25.0 & $1.13\,\pm\,0.20\,$\\
O\,{\scriptsize VII} & 22.10 & 27.3 & $1.11\,\pm\,0.05\,$ & Fe\,{\scriptsize XV} & 52.91 & 2.25 & $1.28\,\pm\,0.21\,$\\
O\,{\scriptsize VIII} & 16.01 & 13.9 & $1.16\,\pm\,0.06\,$ & Fe\,{\scriptsize XV} & 59.40 & 2.99 & $1.38\,\pm\,0.20\,$\\
{\bf O\,{\boldmath \scriptsize{VIII}}} & 18.97 & 88.2 & $1.00\,\pm\,0.02\,$ & {\bf Fe\,{\boldmath \scriptsize{XV}}} & 69.68 & 5.52 & $0.93\,\pm\,0.07\,$\\
Ne\,{\scriptsize VIII} & 88.12 & 5.16 & $1.68\,\pm\,0.29\,$ & Fe\,{\scriptsize XV} & 73.47 & 2.73 & $1.24\,\pm\,0.24\,$\\
{\bf Ne\,{\boldmath \scriptsize{VIII}}} & 98.12 & 1.72 & $1.00\,\pm\,0.20\,$ & {\bf Fe\,{\boldmath \scriptsize{XV}}} & 284.16$^{d}$ &104. & $0.86\,\pm\,0.06\,$\\
{\bf Ne\,{\boldmath \scriptsize{VIII}}} & 98.27 & 4.00 & $1.17\,\pm\,0.13\,$ & Fe\,{\scriptsize XVI} & 50.36 & 3.54 & $2.41\,\pm\,0.24\,$\\
{\bf Ne\,{\boldmath \scriptsize{IX}}} & 13.45 & 22.9 & $1.01\,\pm\,0.05\,$ & Fe\,{\scriptsize XVI} & 50.56 & 1.95 & $2.41\,\pm\,0.27\,$\\
Ne\,{\scriptsize IX} & 13.55 & 7.57 & $1.97\,\pm\,0.16\,$ & Fe\,{\scriptsize XVI} & 54.13 & 2.86 & $2.12\,\pm\,0.27\,$\\
Ne\,{\scriptsize IX} & 13.70 & 16.1 & $1.30\,\pm\,0.07\,$ & Fe\,{\scriptsize XVI} & 54.75 & 4.80 & $1.80\,\pm\,0.17\,$\\
Ne\,{\scriptsize X} & 10.24 & 4.20 & $1.28\,\pm\,0.23\,$ & Fe\,{\scriptsize XVI} & 62.87 & 2.75 & $0.88\,\pm\,0.14\,$\\
{\bf Ne\,{\boldmath \scriptsize{X}}} & 12.14 & 21.5 & $0.92\,\pm\,0.04\,$ & {\bf Fe\,{\boldmath \scriptsize{XVI}}} & 63.71 & 6.25 & $0.97\,\pm\,0.09\,$\\
Mg\,{\scriptsize IX} & 72.31 & $<$1.2 & $<$1.3 & Fe\,{\scriptsize XVI} & 66.25 & 3.81 & $1.76\,\pm\,0.23\,$\\
Mg\,{\scriptsize IX} & 77.74 & $<$0.6 & $<$0.9 & Fe\,{\scriptsize XVI} & 66.38 & 5.96 & $1.84\,\pm\,0.19\,$\\
Mg\,{\scriptsize X} & 57.92 & 2.12 & $0.86\,\pm\,0.19\,$ & {\bf Fe\,{\boldmath \scriptsize{XVI}}} & 335.40$^{d}$ &85.9 & $0.93\,\pm\,0.10\,$\\
{\bf Mg\,{\boldmath \scriptsize{X}}} & 63.31 & 2.82 & $1.02\,\pm\,0.16\,$ & Fe\,{\scriptsize XVI} & 360.75$^{d}$ &45.8 & $1.04\,\pm\,0.10\,$\\
Mg\,{\scriptsize X} & 65.85 & 1.08 & $0.77\,\pm\,0.25\,$ & Fe\,{\scriptsize XVII} & 12.12 & 7.96 & $1.34\,\pm\,0.06\,$\\
Mg\,{\scriptsize XI} & 9.17 & 12.6 & $<$1.7 & Fe\,{\scriptsize XVII} & 12.26 & 7.18 & $1.33\,\pm\,0.15\,$\\
Mg\,{\scriptsize XII} & 8.42 & 3.68 & $1.55\,\pm\,0.28\,$ & Fe\,{\scriptsize XVII} & 13.82 & 3.67 & $0.89\,\pm\,0.10\,$\\
Si\,{\scriptsize X} & 50.52 & 1.87 & $0.76\,\pm\,0.07\,$ & {\bf Fe\,{\boldmath \scriptsize{XVII}}} & 15.02 & 52.1 & $0.98\,\pm\,0.03\,$\\
Si\,{\scriptsize X} & 50.69 & 2.84 & $0.65\,\pm\,0.09\,$ & Fe\,{\scriptsize XVII} & 15.26 & 21.8 & $1.45\,\pm\,0.08\,$\\
Si\,{\scriptsize XI} & 43.76 & 2.90 & $2.13\,\pm\,0.20\,$ & {\bf Fe\,{\boldmath \scriptsize{XVII}}} & 16.78 & 30.4 & $1.02\,\pm\,0.04\,$\\
Si\,{\scriptsize XI} & 49.22 & 3.81 & $1.34\,\pm\,0.10\,$ & {\bf Fe\,{\boldmath \scriptsize{XVII}}} & 17.05 & 77.0 & $1.13\,\pm\,0.06\,$\\
Si\,{\scriptsize XI} & 52.30 & 2.17 & $1.00\,\pm\,0.17\,$ & {\bf Fe\,{\boldmath \scriptsize{XVIII}}} & 14.21 & 10.9 & $1.14\,\pm\,0.10\,$\\
Si\,{\scriptsize XII} & 44.02 & 3.76 & $1.40\,\pm\,0.11\,$ & Fe\,{\scriptsize XVIII} & 16.01 & 3.25 & $1.03\,\pm\,0.06\,$\\
{\bf Si\,{\boldmath \scriptsize{XII}}} & 44.18 & 5.48 & $1.03\,\pm\,0.06\,$ & Fe\,{\scriptsize XVIII} & 16.08 & 5.80 & $1.05\,\pm\,0.12\,$\\
Si\,{\scriptsize XII} & 45.52 & 1.30 & $0.98\,\pm\,0.16\,$ & {\bf Fe\,{\boldmath \scriptsize{XVIII}}} & 93.92 & 7.24 & $1.04\,\pm\,0.07\,$\\
Si\,{\scriptsize XII} & 45.69 & 2.02 & $0.75\,\pm\,0.09\,$ & Fe\,{\scriptsize XVIII} & 103.94 & 2.64 & $1.16\,\pm\,0.18\,$\\
Si\,{\scriptsize XIII} & 6.65 & 8.4 & $<$2.0 & Fe\,{\scriptsize XIX} & 101.55 & 1.47 & $1.15\,\pm\,0.28\,$\\
S\,{\scriptsize XIII} & 35.67 & 4.17 & $1.57\,\pm\,0.21\,$ & Fe\,{\scriptsize XIX} & 108.36 & 2.63 & $0.70\,\pm\,0.11\,$\\
S\,{\scriptsize XIV} & 30.43 & 4.38 & $1.13\,\pm\,0.17\,$ & Fe\,{\scriptsize XX} & 132.84 & $<$4.0 & $<$1.5\\
S\,{\scriptsize XIV} & 32.56 & 3.71 & $0.57\,\pm\,0.09\,$ & Ni\,{\scriptsize XII} & 152.15 & 3.06 & $0.95\,\pm\,0.18\,$\\
Ca\,{\scriptsize XII} & 141.04 &2.74 & $4.24\,\pm\,0.88\,$ & Ni\,{\scriptsize XII} & 154.16 & 1.71 & $1.10\,\pm\,0.32\,$\\
\hline
\renewcommand{\arraystretch}{1}
\end{tabular}
\\
$^a$Lines used in deriving the mean EMD are given in bold face.\\
$^b$Fluxes in $10^{-14}$~erg~cm$^{-2}$~s$^{-1}$, corrected for absorption 
in the ISM.\\
$^c$Predicted from the derived EMD, both using $\log P_{\rm e}=16.10$
(Fig.~\ref{f4} and Table~\ref{tab10}).\\
$^d$Fluxes measured with {\it EUVE} \citep{schmitt96}.\\
$^e$Fluxes measured with {\it FUSE} \citep{simjordan05}.\\
$^f$Fluxes measured with STIS \citep{jordan01a}.
}
\end{flushleft}
\end{table*}

To test how well the EMD of \cite{sf04} reproduces the
fluxes of the EUVE lines of Fe\,{\sc ix} -- Fe\,{\sc xii} and Fe\,{\sc xiv}
 included in our analyses, their volume EMD (that refers to intervals of 
0.10 in $\log T_{\rm e}$) has been converted to the scale of our EMD
over height 
(that refers to intervals of 0.30 in $\log T_{\rm e}$).
The relative abundances derived are also examined for consistency between the
various stages of ionization of a given element. 

The EMD found by \cite{sf04} extends to only $\log T_{\rm e}$ = 5.7 
and so cannot account for the fluxes in the O\,{\sc vi} lines. Our EMD
reproduces all the oxygen resonance lines well and the relative abundances
found from these agree with each other to about a factor of 1.1.
 Similarly, apart from the resonance line of Fe\,{\sc ix} and Fe\,{\sc xiii}
 lines,
our mean EMD leads to the same relative abundance of iron, to within a
factor of 1.3, for all lines used in ions up to and including Fe\,{\sc xviii}.
 \cite{sf04} do not show observed to predicted fluxes for lines of ions
between Fe\,{\sc xi} and Fe\,{\sc xiv}, so we assume that they did not include
lines observed with the {\it EUVE}. We find that using their mean EMD gives
relative iron abundances for these lines that depend on $T_{\rm e}$
and span a 
range of a factor of 2.6. Also, the abundances that we find from 
Ne\,{\sc viii} to {\sc x} are more self-consistent.
Thus, overall, we consider that our mean EMD gives a better representation
of the line fluxes and relative abundances. The reason why we derive similar
mean relative abundances to those found by \cite{sf04} appears
to be the influence of lines used in common in the higher temperature range
where our mean EMDs are in closest agreement.

\cite{wood06} use a similar set of lines to those included by
\cite{sf04}, but adopt CHIANTI (v4.2) for the atomic data. The EMD
that they find peaks at a similar temperature, but below $\log
T_{\rm e} = 6.0$ it is far smaller than ours, probably because they
did not include the O\,{\sc vi} lines observed with {\it FUSE}. 

Since there is interest in the possible presence of a FIP (first ionization
potential) effect, in which elements with a low FIP (less than about 10~eV)
have relatively larger abundances in the corona, we have examined the relative
abundances of oxygen and iron in the corona and photosphere, since this ratio
has the smallest uncertainty in the corona. 
We derive a coronal value of $\log (n_{\rm O}/n_{\rm Fe}) = 1.26 
\pm 0.04$. Combining the photospheric abundances of \cite{asplund05} and 
the differential abundances by \cite{zhao02} or \cite{ap04} leads to
stellar photospheric values of 
$\log (n_{\rm O}/n_{\rm Fe}) = 1.17$ or $1.23$, respectively.
Similarly, adopting the solar photospheric abundances of \cite{grevesse98}
leads to values of 1.29 or 1.35. Thus the largest difference between the 
photospheric and coronal relative abundances is $\pm 0.09$ in the logarithm,
and a larger relative abundance of iron in the corona is found only when
the solar photospheric abundances of \cite{asplund05} are adopted.   
On the basis of 
observations made with the {\it EUVE}, \cite{laming96} concluded that any 
FIP effect in $\epsilon$~Eri was not significantly larger than that found in 
the solar corona. We conclude that there is no clear evidence for {\it any} 
FIP effect in the inner corona of $\epsilon$~Eri. 

There is also considerable interest in the relative abundance of neon to 
oxygen, given its relevance to models of the solar interior \citep{bahcall05}
and the difficulties caused by the lower photospheric abundances of carbon, 
nitrogen, oxygen and neon in reconciling models with the results of 
helioseismology. From Table~\ref{tab8} it can be seen that the coronal value 
derived here is $\log (n_{\rm Ne}/n_{\rm O}) = -0.53$
\citep[c/f -0.44 from \citealt{sf04} or][]{wood06}.
This ratio is significantly larger than the solar values of -0.82 and -0.75 
recommended by \cite{asplund05} and \cite{grevesse98},
respectively. It is also larger than the value of -0.77 found in the solar
transition region by \cite{young06}, and is more similar to the mean stellar
coronal value of -0.39 found by \cite{draketesta05} (who used a more
approximate method to find this ratio). The only obvious
factors that could reduce the derived coronal abundance of neon would be
the present limitations of the current atomic models for Ne\,{\sc viii} and
perhaps Ne\,{\sc ix}. 

Abundances relative to that of iron can also be derived from de-blended lines,
but these do depend on the accuracy of the combined atomic data for the 
lines used. When the Ne\,{\sc x} and Fe\,{\sc xvii} lines at 12.14~\AA\ are 
deblended, the derived flux in the Ne\,{\sc x} line is a factor of 1.09 
smaller than that predicted. If this difference is attributed to an incorrect
neon abundance then the abundance becomes 8.02, not 8.06. The predicted flux
in the Fe\,{\sc xvii} line at 12.29~\AA\ is too small by a factor of
1.33, but we suspect that there are problems with the atomic data since the
EMD based on the adopted abundance of 7.33 gives an overall fit to all the 
iron lines to within a smaller factor. Similarly, when the O\,{\sc viii} and 
Fe\,{\sc xviii} lines at 16.01~\AA\ are deblended, the predicted flux in the
 O\,{\sc viii} line is too small, but only by a factor of 1.16. The predicted
 flux in the Fe\,{\sc xviii} line at 16.07~\AA\ is also too small, but by only
 a factor of 1.05. Thus, to within the expected uncertainties, the relative 
abundances from these individual lines are consistent with those derived 
from the overall fits. 

Table 9 gives the measured line fluxes, corrected for absorption by
the ISM, and the observed to predicted flux ratios for the lines used
in deriving the mean EMD (shown in bold face), based on our abundances given
in Table 8. The flux ratios should be 
interpreted in the light of Section~\ref{S4}, where blending and
atomic data issues are discussed. 

\section{Models based on the energy balance}
\label{S6}

The method used has been set out by \cite{jordanbrown81} for plane parallel
geometry and for a spherically symmetric geometry by \cite{panjordan95}.
It has been extended by \cite{simjordan03a} to include emission from a 
restricted area at a given temperature and is only summarized here. Since 
the heating of the quiet corona occurs at heights much greater than the 
first pressure squared isothermal scale height over which the observed lines 
are predominantly formed, it is assumed that in the regions below, the 
divergence of the conductive flux is balanced by the local radiative losses.
 For a spherically symmetric atmosphere, with emitting area $A_{*}(r)$, and 
in hydrostatic equilibrium, the {\it theoretical} EMD is given by
%\begin{equation}
\begin{eqnarray}
\label{eq8}
\frac{{\rm d}\log [{EM}(0.3)_{\rm th}]}{{\rm d}\log T_{\rm e}} &=&
\frac{3}{2} + 2 \frac{{\rm d}\log P_{\rm e}}{{\rm d}\log T_{\rm e}}
+\frac{{\rm d}\log A_{*}(r)}{{\rm d}\log T_{\rm e}}\\\nonumber
&&-\frac{2 P_{\rm rad}(T_{\rm e}) {EM}(0.3)_{\rm th}^{2}}{\kappa P_{\rm e} P_{\rm H} 
T_{\rm e}^{3/2}}~.\\\nonumber
 % 8
%\end{equation}
\end{eqnarray}
Here, $P_{\rm rad}(T_{\rm e})$ is the radiative power-loss function and $\kappa$ is the
constant in the coefficient of thermal conductivity (the small variations in
$\kappa$ with $T_{\rm e}$ are ignored here). Equation (\ref{eq8}) can be solved
iteratively in hydrostatic equilibrium, to provide the run of the gas and 
electron pressures, the radial height and $EM(0.3)_{\rm th}$ with $T_{\rm e}$. 

We only observe the `apparent' emission measure, given by
\begin{equation}
\label{eq9}
{EM}(0.3)_{\rm app} = {EM}(0.3)_{\rm th} G(r) f(r) \frac{A(r)}{A_{*}(r)}  % 9
\end{equation}
where $f(r) = (r/R_{*})^2$ and $G(r)$ is the fraction of the photons
emitted that are not intercepted by the star, 
\begin{equation}
\label{eq10}
G(r) = 0.5(1 + \sqrt{1 - [1/f(r)]})~.       % 10
\end{equation}
When equation (\ref{eq8}) is used $A_{*}/A(r)$, the fractional area occupied
 at a given r, is set equal to 1.0.

\begin{figure}
\resizebox{\hsize}{!}{\includegraphics{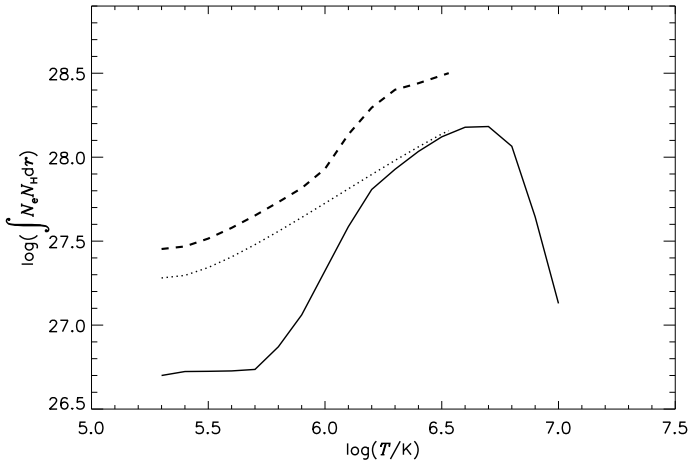}}
\caption{\label{f5}The distribution of $EM(0.3)_{\rm app}$ derived
from the measured line fluxes (full line); $EM(0.3)_{\rm cal}$ derived from
equations 8 and 9, using a spherically symmetric geometry (dotted
line) and no fractional area factor; $EM(0.3)_{\rm cal} A_{*}(r)/A(r)$
(dashed line). The calculated models have $\log T_{\rm c}$ = 6.53; see
also Table 10.}
\end{figure}

\begin{table*}
\begin{flushleft}
\caption{\label{tab10}The apparent and calculated EMDs, the fractional areas
derived, the radial extent above $\log T_{\rm e} = 5.3$ and $P_{\rm e}$ in 
the final theoretical models.}
\begin{tabular}{cc|cccccc}
\hline
$\log T_{\rm e}$ & $\log {EM}(0.3)_{\rm app}^{a}$&$\log {EM}(0.3)_
{\rm cal}
^{b}$ & $[A(r)/A_{*}r]^{c}$ & $\log [{EM}(0.3)_{\rm cal} A_{*}(r)/A(r)]^{d}$ &
 $[A(r)/A_{*}r]^{e}$ & $(r - r_{0})^{f}$ &  $P_{\rm e}$\\
(K) &      &      &      &      &      & $10^{5}$~cm & $10^{16}$~cm$^{-3}$~K \\
\hline
5.30 & 26.700 & 27.281 & 0.26 & 27.453 & 0.18 &  0.0               & 1.38  \\
5.40 & 26.724 & 27.296 & 0.27 & 27.468 & 0.18 &  5.4               & 1.40  \\
5.50 & 26.725 & 27.343 & 0.24 & 27.515 & 0.14 & 1.44 $\times 10$   & 1.41  \\
5.60 & 26.727 & 27.406 & 0.21 & 27.579 & 0.14 & 3.04 $\times 10$   & 1.43  \\
5.70 & 26.736 & 27.479 & 0.18 & 27.653 & 0.12 & 5.98 $\times 10$   & 1.43  \\
5.80 & 26.871 & 27.558 & 0.21 & 27.732 & 0.14 & 1.15 $\times 10^2$ & 1.44  \\
5.90 & 27.061 & 27.641 & 0.26 & 27.816 & 0.18 & 2.19 $\times 10^2$ & 1.44  \\
6.00 & 27.324 & 27.726 & 0.40 & 27.931 & 0.25 & 4.29 $\times 10^2$ & 1.43  \\
6.10 & 27.586 & 27.812 & 0.59 & 28.136 & 0.28 & 9.32 $\times 10^2$ & 1.41  \\
6.20 & 27.808 & 27.897 & 0.81 & 28.295 & 0.33 & 2.16 $\times 10^3$ & 1.37  \\
6.30 & 27.928 & 27.981 & 0.89 & 28.402 & 0.34 & 4.87 $\times 10^3$ & 1.30  \\
6.40 & 28.034 & 28.061 & 0.94 & 28.440 & 0.39 & 1.03 $\times 10^4$ & 1.20  \\
6.50 & 28.121 & 28.138 & 0.96 & 28.487 & 0.43 & 2.09 $\times 10^4$ & 1.07 \\
6.53 & 28.138 & 28.158 & 0.95 & 28.500 & 0.43 & 2.58 $\times 10^4$ & 1.02 \\
\hline
\end{tabular}
\\
$^{a}$ Derived using $\log P_{\rm e} = 16.10$. \\
$^{b}$ Calculated with $A(r)/A_{*}(r) = 1.0$. \\
$^{c}$ Fractional areas calculated from columns 2 and 3. \\ 
$^{d}$ Calculated including a variable area factor.\\
$^{e}$ Fractional areas calculated from columns 2 and 5. \\
$^{f}$ Radial distance above $r = 5.18 \times 10^{10}$~cm at $\log
T_{\rm e} = 5.30$. \\

\end{flushleft}
\end{table*}

In solving equation (\ref{eq8}) it is first assumed that $G(r)=1.0$ and
$f(r)$ = 1.0, and the apparent value of the EMD at a chosen peak coronal
 temperature are used as boundary conditions. The starting value of the total
 gas pressure $P_{\rm g}$ is then found from these parameters and the 
isothermal 
pressure-squared scale height. The values of $G(r)$, $f(r)$ and the starting 
pressure are then updated in each iteration. Thus the iterated solution also 
provides the calculated apparent emission measure $EM(0.3)_{\rm cal}$, which 
can be compared with that observed. The 
further boundary condition applied is that at the base temperature of 
$\log T_{\rm e} = 5.3$, ${\rm d}\log {EM}(0.3)_{\rm app}/{\rm d}\log T_{\rm e}
 = 0$, to fit 
the overall observed mean EMD, including the results from \cite{simjordan05}. 
If the solution is satisfactory, it will also reproduce the electron pressure 
of $15.97\,\pm\,0.2$ at $\log T_{\rm e} = 5.3$. 

Comparing $EM(0.3)_{\rm cal}$ with $EM(0.3)_{\rm app}$ allows any
differences to be attributed to the effects of the emission originating
mainly from a restricted area $A(r)$ at a given $T_{\rm e}$. Additional terms
can then be added to equation (\ref{eq8}) to allow for $A(r)/A_{*}(r)$ and 
its variation with $T_{\rm e}$. This results in eqn. (5) in 
\cite{simjordan03a},
 which gives the full expression for the gradient of 
$\log [EM(0.3)_{\rm cal} A_{*}(r)/A(r)]$ with $\log T_{\rm e}$. The values of 
$A(r)/A_{*}(r)$ can then be refined.  
 
There are two differences between the application of the code used here and
by \cite{simjordan03a}. First, the maximum EMD is always associated with
an isothermal region at the chosen coronal temperature, secondly, the form
of $P_{\rm rad}(T_{\rm e})$ adopted is $2.80 \times 10^{-19}/T_{\rm e}^{1/2}$,
 to take account of the additional lines now included in atomic data bases.  

After investigating a number of models using equation (\ref{eq8}), the one
that gives the best fit to the constraints set out above has a coronal 
temperature of $\log T_{\rm e} = 6.53$, a peak emission measure of 
$\log EM(0.3)_{\rm cal} = 28.16$ and a base pressure of 
$\log P_{\rm e} = 15.97$. The EMD derived from the line fluxes and
this solution for $EM(0.3)_{\rm cal}$ are
shown in Fig.~\ref{f5} by full and dotted lines, respectively. The area 
factors derived are given in column 4 of Table~\ref{tab10}. 

Because of the form of the energy balance equation and the boundary 
conditions chosen, the choice of the coronal 
temperature determines the coronal emission measure and pressure.
The ratio of the base pressure (at $\log T_{\rm e} = 5.3$) to the
coronal pressure 
has a constant value and the base pressure and temperature determine the
theoretical value of the base $EM(0.3)$, and hence the fractional area at
the base temperature. Thus the parameter space of the models can be further
explored without additional full calculations. The scaling laws that result
will be discussed by Jordan \& Ness (in preparation). 

The apparent EMD is poorly determined between 
$\log T_{\rm e} = 5.3$ and 5.8, owing to the paucity of lines observed
in this 
temperature range. When the derived areas are used in the full
equation (5) of 
\cite{simjordan03a}, the area is kept constant from $\log T_{\rm e} =
5.3$ to 5.8, 
at the value at $\log T_{\rm e} = 5.5$, where the lines of O\,{\sc vi} are 
predominantly formed. This is justified by the small physical extent of this
region.
The resulting distribution of $\log [EM(0.3)_{\rm cal} A_{*}(r)/A(r)]$
as a function of $\log T_{\rm e}$
 is shown in Fig.~\ref{f5}, and the new area factors that result are
given in
column 6 of Table~\ref{tab10}. The radial extent and electron pressure
in the final model are also given in Table~\ref{tab10}. A further
iteration with the new area factors 
was carried out to check that the new solution did not differ significantly 
from the previous one, but given the inherent uncertainties in the fluxes, 
atomic data and the constants used in the energy balance equation, a fully 
converged solution was not pursued. 

The mean EMD derived directly from the observed line fluxes peaks at a 
temperature around $\log T_{\rm e} = 6.6\,\pm\,0.05$ and the lines of 
Fe\,{\sc xviii}
and to a lesser extent, Fe\,{\sc xvii}, do appear to require material at 
higher temperatures. We attribute this 
emission to active regions, but cannot model them to remove their 
contribution, since the electron pressure is not known, and the energy balance
used here will not be appropriate, since heating by other than thermal 
conduction is not included. Quiet coronal models with 
$\log T_{\rm e} \ge 6.6$ that satisfy the constraint on the emission measure 
gradient at $\log T_{\rm e} = 5.3$ lead to base values of $\log P_{\rm e}$ 
that exceed  
 the upper limit of 16.17 derived from the density-sensitive lines at about 
this temperature.

It is difficult to make detailed comparisons with other earlier work, but
there are several early determinations of the temperature at which the peak
EMD occurs. 
\cite{giampapa85} used observations of $\epsilon$~Eri made with the 
Imaging Proportional Counter (IPC) on the {\it Einstein Observatory} to deduce
a single-temperature fit of $\log T_{\rm e} = 6.53$, close to the
value of 6.60 ($\pm 0.05$) at which we and \cite{wood06} find the EMD to peak.
 However, the loop
models that \cite{giampapa85} investigated had pressures that exceeded those 
found here and 
they remark that they could not simultaneously explain the IUE and X-ray 
spectra. \cite{schmitt90} also used these observations but could not 
find a single-temperature fit to the spectrum. The two-temperature fit that 
they suggested is not consistent with the present results. Because there was
some suggestion that the absence of stars with coronal temperature
($T_{\rm c}$) between
$\log T_{\rm e} = 6.67$ and 6.88 in the sample studied by \cite{schmitt90},  
might arise from the energy response function of the IPC, \cite{montjor93}
used their scaling law between $T_{\rm c}$, $g_*$ and the Rossby number, 
$Ro$, to predict the coronal temperature for $\epsilon$~Eri. With the 
currently adopted value of $g_*$, the scaling law predicts 
$\log T_{\rm c} = 6.57$, close to the value found here. The emission measures
shown by \cite{laming96}, based on observations with the {\it EUVE},
also peak at $\log T_{\rm e} = 6.5\,\pm\,0.1$. Thus there is good
agreement between
the peak temperature from the observations made with three different
instruments.   

The absolute scale of $EM(0.3)_{\rm app}$ that we adopt depends on the value
of the iron abundance used. If a smaller value were used, the values of
$EM(0.3)_{\rm app}$ would all increase. The intercept with $EM(0.3)_{\rm cal}$
would occur at a {\it lower} temperature and above this value, the area factors
would exceed 1.0, which is not physically acceptable. Thus it seems unlikely
that the adopted value of the iron abundance is significantly too large.
Conversely, using a larger iron abundance would lead to smaller values of
$EM(0.3)_{\rm app}$. This would result in smaller values of $A(r)/A_{*}(r)$
by the same factor. Such solutions cannot be excluded.
      
\section{Discussion and conclusions}
\label{S7}

We have analysed line fluxes measured with the LETGS on {\it Chandra} to
obtain an apparent EMD. As part of this work we have examined the
self-consistency of the results from lines of a given ion. There remain
inconsistencies in the results from different lines of Fe\,{\sc xvi} and,
 to a lesser extent, Fe\,{\sc xv}. One source of these
differences could be excitations to n-states not yet included in the
atomic models, followed by cascades. 
 Although the lines are weak, there is a significant
 difference between the observed and predicted fluxes for the 3p -- 2s 
transition in Ne\,{\sc viii}. The atomic models and data do not appear 
to have been updated in CHIANTI since v3. There also needs to be a proper
treatment of recombination (including cascades) to the 1s2s~$^3$S level 
in the He\,{\sc i}-like ions and di-electronic recombination needs to be 
included. The pressure indicated by the f/i ratio in O\,{\sc vii} is
currently somewhat larger than expected from the final model. Although
blends have been taken into account in
analysing the Fe\,{\sc xiii} lines at around 203.8~\AA\, the observed flux
is lower than predicted by CHIANTI (v5.2), unless $\log P_{\rm e}$ is
lower than 15.30, which is not consistent with the results from other
lines.

Line optical depths can be estimated using the final model that includes
the variable area factors. Several lines, in particular the resonance
line of Fe\,{\sc ix} at 171~\AA, are estimated to have line-centre
optical depths approaching 1. The effects on the measured fluxes
will depend on the geometry; scattering {\it out of} the line of sight
would be expected for lines formed in supergranulation boundaries, but
to find the expected flux when integrated over the whole star would
require detailed radiative transfer calculations.               

Relative element abundances have been determined in the upper transition 
region/corona. These agree with those found previously by \cite{sf04}
to within the expected uncertainties, and quite well with those of 
\cite{wood06}, in spite of differences in the 
mean EMDs. This reproducibility lends support to the individual line-based 
methods of deriving abundances. The EMD found here is based on line fluxes 
measured from the
LETGS spectrum, the STIS spectrum (for O\,{\sc vi}) and on the {\it EUVE} 
counts measured by \cite{schmitt96} for Fe\,{\sc ix} to {\sc xii} and 
Fe\,{\sc xiv} to {\sc xvi}. This EMD
gives a consistent relative element abundance of iron for all stages of
 ionization included, apart from Fe\,{\sc ix}, for which the line flux is 
observed to be weaker than predicted. Using the EMD
found by \cite{sf04} leads to a larger difference between the abundances
found from Fe\,{\sc xi}, {\sc xii} and {\sc xiv}. On the basis of the 
relative abundances of oxygen and iron, we conclude that
there is no clear evidence of
any difference between photospheric and coronal abundances for low and high
FIP elements. Indeed, none of our previous studies of stellar
transition regions has shown any clear evidence of FIP effects. The 
spatially integrated X-ray line fluxes are dominated by the first
pressure-squared isothermal scale height; if changes in relative
element abundances are occurring at greater heights in the corona, they might
not be detectable. (See also discussion by \citealt{wood06} regarding 
correlations with stellar mass-loss rates.)  

The important Ne/O relative abundance is found to be -0.53 (on a logarithmic
scale), slightly smaller than the value of -0.44 found by \cite{sf04}
and \cite{wood06}, but larger than the values of -0.82 and -0.75
recommended by 
\cite{asplund05} and \cite{grevesse98}, respectively, for the solar 
atmosphere. It is also larger than that found for the solar 
transition region by \cite{young06} (-0.77). \cite{draketesta05}
derived an even larger value (-0.39) from studies of a range of
active stars, but used a more approximate method, not a full study of the EMD.
Since the relative abundances of O and Fe found in the stellar corona
agree with the range of stellar photospheric values to within $\pm
0.09$ in the logarithm, it appears that it is the abundance of Ne in
the stellar corona that differs from the recommended solar values.    
\cite{draketesta05} have suggested that a larger Ne/O abundance ratio
could resolve the difficulties introduced by the adoption of the lower C,
N, and O abundances proposed by \cite{asplund05}.

The similarity of the variation of the EMD with $T_e$ found by using
lines from individual isoelectronic sequences or just one element can be 
seen from Figs. 3 and 4. We are of the opinion that line-based analyses,
using the full emission measure contribution functions is the best method 
for determining the mean EMD. 
     
Recent work by the other authors mentioned above has concentrated on deriving
a mean EMD and element abundances but models of the atmosphere were not
produced.  
We regard the main purpose of deriving the mean EMD is to use it in 
comparisons with 
theoretical models based on assumptions about the energy balance. Early work 
on modelling in terms of loop structures was not entirely
successful \citep[e.g. that by][]{giampapa85}. We intend to
study a larger sample of stars in future work, to investigate the
systematic behaviour with stellar activity. 
  
The mean (apparent) EMD has been compared with the predictions of models based
on the assumption of an energy balance between the divergence of the thermal
conductive flux and the radiation losses. This allows the fractional area
of the emitting material to be found as a function of $T_{\rm e}$. When the 
variation of the fractional area is not included, we find a fractional area 
that is constant at about 0.24 up to $\log T_{\rm e} = 5.8$ and then 
increases with $T_{\rm e}$ to reach about 1 by $\log T_{\rm e} =
6.4$. Allowing for the variation of
the fractional area with $T_{\rm e}$ increases the calculated values of 
$P_{\rm e}$ and reduces the base fractional to 0.14 and the 
coronal area factor to 0.43. The solutions derived here do not have
the problem of fractional areas that are greater than 1 that occurred
in the earlier analyses by \cite{simjordan05}.   
The derived behaviour is similar to the trend in the area occupied by the 
supergranulation network boundaries in the solar transition region and inner 
corona, taking into account that the stellar transition region extends to
higher temperatures, because the coronal temperature is higher.   
At the lower end of the temperature range, the surrounding material would be
at near coronal temperatures; the coronal filling factor of less than
one is consistent with the additional presence of active regions.

\section*{Acknowledgments}

J.-U.N acknowledges support from PPARC under grant number PPA/G/S/2003/00091
and from NASA through {\it Chandra}
Postdoctoral Fellowship grant PF5-60039 awarded by the {\it Chandra} X-ray
Center, which is operated by the Smithsonian Astrophysical Observatory for
NASA under contract NAS8-03060. We thank the referee for useful
comments on photospheric abundances. 

\bibliographystyle{apj}
\bibliography{mybibfile}

\label{lastpage}

\end{document}